\documentclass[12pt,a4paper,epsf]{article}
\usepackage{graphics}
\usepackage{amssymb,amsmath}
\usepackage[dvips]{lscape,graphicx}
\usepackage{cite}
\usepackage{longtable}
\usepackage{slashed}

\newcommand{\ct}{\cite}
\newcommand{\lb}{\label}

\newcommand{\bc}{\begin{center}}
\newcommand{\ec}{\end{center}}
\newcommand{\bd}{\begin{displaymath}}
\newcommand{\ed}{\end{displaymath}}
\newcommand{\be}{\begin{equation}}
\newcommand{\ee}{\end{equation}}
\newcommand{\ba}{\begin{array}}
\newcommand{\ea}{\end{array}}
\newcommand{\bea}{\begin{eqnarray}}
\newcommand{\eea}{\end{eqnarray}}
\newcommand{\bt}{\begin{tabular}}
\newcommand{\et}{\end{tabular}}
\newcommand{\un}{\underline}

\newcommand{\bp}{\begin{picture}}
\newcommand{\ep}{\end{picture}}
\newcommand{\bfi}{\begin{figure}}
\newcommand{\efi}{\end{figure}}

\def\fun#1#2{\lower3.6pt\vbox{\baselineskip0pt\lineskip.9pt
\ialign{$\mathsurround=0pt#1\hfil##\hfil$\crcr#2\crcr\sim\crcr}}}

\begin{document}

\title{\LARGE \bf {Planck Scale Physics, Gravi-Weak Unification and the Higgs Inflation}}
\author{\large \bf  L.V. Laperashvili,${}^{1}$\footnote
{laper@itep.ru,}\,\, H.B. Nielsen ${}^{2}$\footnote{hbech@nbi.dk}
and B.G. Sidharth ${}^{3}$\footnote{iiamisbgs@yahoo.co.in}\\[5mm]
{\large \it ${}^{1}$ The Institute of Theoretical and
Experimental Physics,}\\
{\large\it National Research Center ``Kurchatov Institute'',}\\
{\large\it Bolshaya Cheremushkinskaya, 25, 117218 Moscow, Russia}\\
{\large \it ${}^{2}$ Niels Bohr Institute,}\\
{\large \it Blegdamsvej, 17-21, DK 2100 Copenhagen, Denmark}\\
{\large \it and}\\
{\large \it ${}^{3}$ International Institute of Applicable Mathematics
and Information Sciences,}\\
{\large \it B.M. Birla Science Centre}\\
{\large \it Adarsh Nagar, Hyderabad - 500063, India}}
\date{}
\maketitle

\thispagestyle{empty}

\vspace{2cm}

{\bf Keywords:} gravity, unification, mirror world, cosmological constant,
dark energy, inflation, connection, tetrads.

{\bf PACS:}  04.50.Kd,  98.80.Cq,
12.10.-g, 95.35.+d, 95.36.+x

\clearpage \newpage

\begin{abstract}

Starting with a theory of the discrete space-time at the Planck
scale, we developed a Gravi-Weak Unification (GWU) - a
$Spin(4,4)$-invariant model unified gravity with weak $SU(2)$
gauge and Higgs fields in the visible and invisible sectors of the
Universe. Considering the Gravi-Weak symmetry breaking, we showed
that the obtained sub-algebras contain the self-dual left-handed
gravity in the OW, and the anti-self-dual right-handed gravity in
the MW. Finally, at the low energy limit, we have only the
Standard Model (SM) and the Einstein-Hilbert's gravity. The
Froggatt-Nielsen's prediction of the top-quark and Higgs masses
was given in the assumption that there exist two degenerate vacua
in the SM. This prediction was improved by the next order
calculations. We have developed a model of the Higgs Inflation
using the GWU action. According to this inflationary model, a
scalar field (inflaton) starts trapped from the "false vacuum" of
the Universe at the Higgs field's VEV $v_2 \sim 10^{18}$ GeV. The
interaction between the ordinary and mirror Higgs fields $\phi$
and $\widetilde{\phi}$ generates a Hybrid model by A.~Linde of the
Higgs Inflation in our Universe.

\end{abstract}

\vspace{1cm}

{\large \bf Contents:}\\

{\bf

1. Introduction.\\

2. Discrete space-time of the Universe at the Planck scale.\\

~~~~a) Sidharth's theory of the discrete space-time.

~~~~b) Random Dynamics and Multiple Point Principle.\\

3. Plebanski's formulation of General Relativity.\\

4. Gravi-Weak Unification in the visible sector of the Universe.\\

5. Mirror world with broken mirror parity.\\

~~~~a) Communications between visible and hidden worlds.

~~~~b) Gravi-Weak action in the invisible (mirror) sector of the
Universe.\\

6. U(4)-group of fermions.\\

7. Multiple Point Model and the prediction of the top and Higgs
Masses.\\

8. The Higgs Inflation model with the mirror Higgs boson. \\

~~~~a) The GWU action and the self-consistent Higgs Inflation
model.\\

9. Summary and Conclusions.}

\clearpage \newpage

\section{Introduction}

In this investigation we start with a theory of the discrete
space-time by B.G. Sidharth \ct{1S,2S,3S,4S}, existing at the very
small (Planck length) distances. Using the results of
non-commutativity, the author has shown that in our Universe the
cosmological constant (or dark energy density) is almost zero in
initial stage of the Universe and so far.

Random Dynamics (RD), suggested by H.B. Nielsen \ct{1R,2R} and his
collaborators \ct{3R,4R,5R,6R,7R,8R}, is also related with a
discrete space-time. RD assumes that at the very small Planck
scale distances the fundamental law of the Nature is randomly
chosen from a large set of different theories. We assume that such
an initial theory, randomly chosen at the Planck scale, is a
theory with a group of symmetry:
\be G = G_{(GW)} \times U(4),\lb{e1} \ee
where $G_{(GW)}$ is a group of Gravi-Weak Unification:
\be G_{(GW)} = SO(4,4) \sim Spin (4,4), \lb{e2} \ee
and $U(4)$ is a group of fermions. Previously gravi-weak and
gravi-electro-weak unified models were suggested in
Ref.~\ct{1gw,2gw,3gw}. The gravi-GUT unification was developed in
\ct{1Lizi,1gu,2gu,3gu}. In Ref.~\ct{1u} a model of unification of
gravity with the weak $SU(2)$ gauge and Higgs fields was
constructed, in accordance with Ref.\ct{2u}. However, in contrast
to Ref.~\ct{2u}, we assume the existence of the hidden sector of
the Universe, when this hidden world is a Mirror World (MW) with
broken Mirror Parity (MP) (see Refs. \ct{1mw,2mw,3mw,4mw,5mw} and
reviews \ct{6mw,7mw,8mw,9mw}). In the present paper we give
arguments that MW is not identical to the visible Ordinary World
(OW). We consider an extended $\mathfrak g =
\mathfrak{spin}(4,4)_L$-invariant Plebanski action in the visible
Universe, and $\mathfrak {g = spin}(4,4)_R$-invariant Plebanski
action in the MW. Then we show that the Gravi-Weak Unification
symmetry breaking leads to the following sub-algebras: $
{\mathfrak g}_1 = {\mathfrak {sl}}(2,C)^{(grav)}_L \oplus
{\mathfrak {su}}(2)_L$ -- in the ordinary world, and $\widetilde
{{\mathfrak g}_1} = {\widetilde {\mathfrak {sl}}}(2,C)^{(grav)}_R
\oplus {\widetilde {\mathfrak {su}}}(2)_R$ -- in the hidden world.
These sub-algebras contain the self-dual left-handed gravity in
the OW, and the anti-self-dual right-handed gravity in the MW.
Finally, at low energies, we obtain the Standard Model (SM) group
of symmetry and the Einstein-Hilbert's gravity. In this approach
we construct a model of the Inflation, in which the inflaton
$\sigma$, being a scalar field, during inflation decays into the
two Higgs doublets of the SM: $\sigma \to \phi^\dagger \phi$, and
then the interaction between the ordinary and mirror Higgs boson
fields develops the Hybrid model of the Inflation \ct{Lin}.

\section{Discrete space-time of the Universe at the Planck scale}

At present time the physicists are keenly and persistently
searching for some fundamental theory at the Planck scale, at very
small distances, much smaller than we can study directly
experimentally (with present accelerators). Trying to look insight
the Nature and considering the physical processes at small
distances, physicists have made attempts to explain the well known
low-energy Standard Model (SM) results as a consequence of the
Planck scale physics.

Here we have two possibilities:\\
1) At the very small (Planck length) distances our space-time is
continuous and there exists
a fundamental theory maybe with a very high symmetry.\\
2) At the very small distances our space-time is discrete, and
this discreteness influences
on the Planck scale physics and beyond it.\\
In this talk we review, discuss and classify  some attempts
(exactly, only few examples) to guess such a fundamental theory,
which \un{could} be the fundamental Theory of the Everything
(TOE).

We shall first look at how we might classify such theories.

It is one very important classification separation: whether in the
space-time we have all points (continuum), or we have only certain
points, rather than all of them. Here, of course, we refer to
lattice-like theories as the ones with only some points of space
existing, while the continuum space-time represents all Grand
Unification Theories (GUTs), superstring theory, General
Relativity (GR), supersymmetry, etc. As an example of a continuum
space-time having all points, can be considered a theory with the
Euclidean geometry in the modern axiomatic formulation (Hilbert's
axiomatic representation).

There are many ways to vary in details of how a theory could have
only discrete points. The assumption that $(3+1)$- dimensional
space-time is discrete on the fundamental level is an initial
(basic) point of view of the present investigation: we take the
discreteness as existing, not as the lattice computation trick (as
in QCD, say). In the simplest case we can imagine our $(3+1)$
space-time as a regular hypercubic lattice with a parameter
$\lambda_{Pl}$, where $\lambda_{Pl}$ is the Planck length. But of
course we do not know (at least, on the level of our today
knowledge), what lattice-like structure (random lattice, or
graphene, or any crystal, or foam, or string lattice, etc.) is
realized in the description of physical processes at the very
small distances.

Modern Fuzzy Space-time models, Loop Quantum Gravity (see for
example \ct{Rov}) and a few other approaches start
from the Planck scale as a minimum scale of the Universe.\\
This is also the starting point of the alternative theory, which
was developed by B.G. Sidharth in his book \ct{1S} and in the
large number of his papers (see Refs.~ \ct{2S,3S,4S}). It has been
considered that the space-time is fuzzy, more generally
non-differentiable, and is presented by a non-commutative
geometry. The concept of non-commutativity leads to an essential
predictions, described in the book \ct{1S} and used in the present
paper.

Random Dynamics (RD), suggested and developed by H.B. Nielsen and
his collaborators \ct{1R,2R,3R,4R,5R,6R,7R,8R}, also is a theory
of physical processes proceeding at small distances of order of
the Planck length. RD tries to derive the laws of physics known
today by the use of almost no assumptions.

Previously the efforts to construct a fundamental theory had led
to the Grand Unified Theories (GUTs), especially Supersymmetric
SUSY GUTs, which had an aim to give an unified self-consistent
description of electroweak and strong interactions by one simple group:
$SU(5), SU(6), SO(10), E_6, E_8,$ etc.\\
The group of Theory of the Everything (TOE) was suggested in
Refs.~\ct{1Lizi}.
But at present time the experiment does not indicate any manifestation of these theories.\\
The next step to search the fundamental theory was a set of
theories describing the extended objects: strings, superstrings
and $M$-theories. They came into existence
due to the necessity of unification of electroweak and strong
interactions with gravity.\\
RD is an alternative to these theories. The above mentioned
theories are based on some fixed axioms. If namely one of these
axioms is changed a little bit, then the theory also is changed
and often becomes not even consistent. In RD we have the opposite
case. RD is based on the very general assumptions, which take
place at the fundamental scale.

Theory of Scale Relativity (SR) \cite{1N} is also related with a
discrete space-time. This theory assumes that the resolution of
experimental measurements plays in quantum mechanics an important
role. Suggesting to consider the non-differentiable space-time, L.
Nottale built the microphysical world on the concept of the
fractal space-time. This non-differentiability implies an explicit
dependence of space-time on scale. The principle of relativity was
applied not only to motion, but also to scale transformation,
because the resolution measurements must be taken into account in
definition of the coordinate system.\\
The resolution of the measurement apparatus plays in
quantum physics a completely new role
with respect to the classical, since the result of measurements
depends on it as a consequence of Heisenberg's relations.\\
The resolution characterizes the system under consideration. Complete
information about the measurement
position and time can be obtained, when not only
space-time coordinate $(t,x,y,z)$, but also
resolutions $(\Delta t, \Delta x, \Delta y, \Delta z)$ are
taken into account. As a result,
we expect that the space-time to be described by a metric:
$$g_{\mu \nu} = g_{\mu \nu} (t,x,y,z ; \Delta t, \Delta x, \Delta y, \Delta z).$$

In connection with a theory of the discrete space-time, we can
mention the model by H. Kleinert \ct{1k}, in which he has shown
that there exists a close analogy of geometry of space-time in GR
with a structure of a crystal with defects. Kleinert's model
considers the translational defects -- dislocations, and the
rotational defects -- disclinations -- in the 3- and 4-dimensional
crystals. Ref.\ct{3k} presents the relation between the Kleinert's
model of a crystal and the Plebanski's formulation of gravity. It
was shown that the tetrads in gravitational theory are dislocation
gauge fields, and the connections are the disclination ones, the
curvature is the field strength of connections, and the torsion is
the field strength of tetrads.

In this investigation we start with a theory of the discrete
space-time by B.G. Sidharth \ct{1S}.

\subsection{Sidharth's theory of the discrete space-time and the predictions
of non-commutativity}

The concept of the discrete space-time (item 2) is a base of the
theory, which was developed by B.G. Sidharth in his book \cite{1S}
and in his papers, for example, in Refs.~\ct{2S,3S,4S}.

B.G. Sidharth was first (1997 year) - before the discovery of S.
Perlmutter's et al. \ct{Per} in 1998 year, - who has shown that a
cosmological constant is very small:
$$\Lambda \sim H_0^2,$$
where $H_0$ is the Hubble rate. Then the Dark Energy (DE) density
is also very small: $$\rho_{DE}\sim 10^{-48}\,\,{\rm{GeV}}^4,$$
what provided the accelerating expansion of our Universe after the Big Bang.\\
B.G. Sidharth deduced the cosmological constant and dark energy density from
the following points of view.\\
Modern Quantum Gravity (Loop Quantum Gravity \ct{Rov}, etc.,) deal
with a non-differentiable space-time manifold. In such an approach
there exists a minimal space-time cut off, which leads to the
non-commutative geometry, a feature shared by
the Fuzzy Space-Time also \ct{1S,1F,2F,3F,4F}.\\
A starting point of the book \ct{1S} is the well known fact that
in random walk, the average distance $l$ covered at a stretch is
given by
\be l = R / \sqrt{N}, \lb{1s} \ee
where $R$ is the dimension of the system and $N$ is the total number of steps.\\
The Eq. (\ref{1s}) is true in the Universe itself with $R_{un}$ being the radius of
the Universe $\sim 10^{28}cm$; $N_{un}$ is the number of elementary particles in the Universe
$(N_{un} \sim 10^{80})$, and $l$ is the Compton wavelength of the typical elementary particle
with mass $m$ $(l = m/\hbar c)$ ($l\sim 10^{-10}cm$ for electron).\\
A similar equation for the Compton time exists in terms of the age $(T_{un})$ of the Universe:
\be
T_{un} = \sqrt{N_{un}} \tau ,\lb{2s}
\ee
where $\tau$ is a minimal time interval (chronon).\\
Eqs.(\ref{1s}) and (\ref{2s}) arise quite naturally in a cosmological scheme
based on fluctuations.\\
If we imagine that the Universe is a collection of the Planck mass oscillators,
then the number of these oscillators is:
\be
        {N}_{un}^{Pl}\sim 10^{120}. \lb{3s}
\ee If the space-time is fuzzy, more generally non-differentiable,
then it has to be described by a non-commutative geometry with the
coordinates obeying the following commutation relations:
\be [dx^\mu , dx^\nu ] \approx \beta^{\mu \nu} l^2 \ne 0.\lb{4s}
\ee
Eq. (\ref{4s}) is true for any minimal cut off $l$.\\
Previously the following commutation relation was considered by
H.S. Snyder \ct{1F}:
\be [x,p] = \hbar [1 + (\frac{l}{\hbar})^2 p^2],\,\, etc., \lb{5s}
\ee
which shows that effectively 4-momentum $p$ is replaced by
\be p \to p (1 + \frac{l^2}{\hbar^2} p^2)^{-1},\lb{6s} \ee
and the energy-momentum formula now becomes as:
\be E^2 = m^2 + p^2 (1 + \frac{l^2}{\hbar^2} p^2)^{-2},\lb{7s} \ee
or
\be E^2 \approx m^2 + p^2 - \gamma \frac{l^2}{\hbar^2} p^4,\lb{8s}
\ee
where $\gamma \sim 2$.\\
In such a theory the usual energy momentum dispersion relations are
modified \ct{3S,4S}.\\
In the above equations $l$ stands for a minimal (fundamental)
length, which could be the Planck length, or more generally
Compton wavelength. It is necessary to note, that if we neglect
order of $l^2$ terms, then we return to the usual quantum theory,
or quantum field theory. Writing Eq.~(\ref{8s}) as
\be E = E' - E'',  \lb{9s} \ee
where $E'$ is the usual (old) expression for energy, and $E''$ is
the new additional term in modification \ct{4S}, $E''$ can be
easily verified as
\be E'' = mc^2. \lb{10s} \ee
In Eq. (\ref{10s}) the mass $m$ is the mass of the field of bosons.\\
Furthermore it was shown, that (\ref{8s}) is valid only for boson
fields, whereas for fermions the extra term comes with a positive
sign \ct{2S,3S,4S}. In general, we can write:
\be E = E' + E'' , \lb{11s} \ee
where $E'' = -m_b c^2$ -- for boson fields, and $E'' = + m_f c^2$
-- for fermion fields (with mass $m_b, m_f$, respectively).

These formulas help to identify the DE density, what was first
realized by B.G. Sidharth in Ref.\ct{2S}.

DE density is the density of the quantum vacuum energy of the
Universe. Quantum vacuum, described by Zero Point Fields (ZPF)
contributions, is the lowest state of any Quantum Field Theory
(QFT), and due to Heisenberg's principle has an infinite value,
which is "renormalizable".

Quantum vacuum of the Universe can be a source of cosmic repulsion
(see \ct{1Z} and \ct{10S}). However, a difficulty in this approach
has been that the value of the cosmological constant turns out to
be huge, far beyond what is observed by astrophysical
measurements. This has been
called "the cosmological constant problem" \ct{1W}.\\
The mysterious ZPF contributions, or quantum vacuum energy density
(dark energy DE), has been experimentally confirmed by
astrophysical measurements. It is given by the following value
\ct{1D,2D,3D}:
\be \rho_{DE} \approx (2.3 \times 10^{-3}eV)^4. \lb{12s} \ee
The value of cosmological constant $\Lambda$ is related with
$\rho_{DE}$ by the following way:
\be \Lambda = 8 \pi G_N \rho_{DE} = \rho_{DE}/(M^{(red.)}_{Pl})^2,
\lb{13s} \ee
where $G_N$ is the Newton's gravitational constant, and
$M^{red.}_{Pl}$ is the reduced Planck mass.

Using the non-commutative theory of the discrete space-time, B.G.
Sidharth predicted in \ct{1S,2S} the value of cosmological
constant $\Lambda$:
\be \Lambda  \simeq H_0^2, \lb{14s} \ee
where $H_0$ is the Hubble rate (see \ct{1D,2D,3D}):
\be   H_0 \simeq 1.5\times 10^{-42}\,\,{\rm{GeV}}. \lb{15s} \ee
The very interesting predictions of the non-commutativity lead to
the physically meaningful relations, including a rationale for the
Dirac equation
and the underlying Clifford algebra (see \ct{10S}).\\
B.G. Sidharth also predicted the light neutrino mass and tried to
extract cosmological constant from the Fermi energy of the cold
primordial neutrino background \ct{11S}.

In connection with a theory of the non-commutativity, it is useful
to see \ct{CNT}.

We have essentially used the results of \ct{1S} in our Higgs
inflation model.

\subsection{Random Dynamics and Multiple Point Principle}

Random Dynamics (RD) was suggested and developed in
Refs.~\cite{1R,2R,3R,4R,5R,6R,7R} (see also a review \ct{8R}) as a
theory of physical processes proceeding
at small distances of order of the Planck length.\\
In Random Dynamics dream (speculation) it is hoped that the
physics at very small distances is almost unimportant for what the
effective laws to be observed by physicists working with energies
per particle up to only a few TeV. So one could crudely say:
Random Dynamics does NOT know the physics at small distances. It
also does not matter is the hope. But if one has in mind that the
true mathematical definition of the even uncountable set of real
numbers is a bit tricky, one might say that we can hardly believe
that the most fundamental physics should be based on manifolds
truly.

Based on the very general assumptions at the fundamental level,
the RD argues: the fundamental laws of the Nature are so
complicated that it is preferable to think that at very small
distances there are no laws at all, our space-time is discrete and
the physical processes are described randomly. Then the
fundamental law of the Nature is one, which was randomly chosen
from a large set of sufficiently complicated theories, and this
one leads to the laws observed in the low-energy limit of our
experiment.

The lattice model of gauge theories is the most convenient
formalism for the realization of the RD ideas.

In lattice Monte Carlo calculations with lattice theories with
several action terms coming with each its coefficient (coupling),
say traces of different representations of the plaquette variables
$U(\Box)$, one finds phase transitions and even say a triple
point. There can then even be phases some invariant subgroup such
as say $Z_N$ of $SU(N)$ could have confinement at the lattice
scale, while the full group might first be confining at a much
longer space scale.

Long ago \ct{1M} (see also \ct{2M}) it was used the assumption
that the coefficients in the lattice action were just being
adjusted to be at the triple or higher point - Multiple Critical
Point (MCP) - in the phase diagram of a theory, to fit
finestructure constants in the Antigrand model \ct{1AG,2AG}. The
Anti-GUT model, in which each family of quarks and leptons have
their own Standard Model group were inspired by some random
dynamics thinking.

The a priori mysterious assumption that the coefficients in the
lattice gauge theory action should be just at the critical point,
where several phases meet, gave rise to that D.L. Bennett and H.B.
Nielsen \ct{1M} invented the idea of "Multiple Point Principle"\,
(MPP), which claims that coupling constants get adjusted to bring
the vacuum into multiple point where several phases, several types
of vacua, can coexist in the sense of having the same energy
density.

At first, this MPP seems not to be a consequence of random
dynamics thinking, but by a somewhat round about argumentation one
can however claim that indeed random dynamics leads to the
multiple point principle, which postulates: all vacua which might
exist in the Nature (as minima of the effective potential) should
have zero, or approximately zero cosmological constant.

One relatively recent claim in the RD is that even if the action
were complex, rather than real as usually assumed, it would
approximately not be seen in the usual equations of motion and
only some predictions about the initial state, or rather about
what really happens, would reveal the imaginary part of the action
(or equivalently the anti-Hermitean part of the Hamiltonian). If
one use this result to assume that there fundamentally is an
imaginary part also of the action, this leads to an extremization
principle that this imaginary part of the action should be
minimized - as may be crudely seen by thinking of the
Wentzel-Dirac-Feynman path way integral, in which the integrand
goes exponentially down numerically with increasing imaginary part
of the action. If it were even speculated, that somehow the
coupling constant in the effective theory could also be varied in
the search for minimal imaginary part of the action, then it could
very easily happen that we got several degenerate vacua, i.e. we
could then very likely get the multiple point principle \ct{7R}.

Finally, we conclude that the Multiple Point Model (MPM)
\ct{1M,2M} postulates:\\ {\it there are many vacua in the Universe
with the same energy density, or cosmological constant, and all
cosmological constants are zero, or approximately zero}.\\
For example, N. Arkani-Hamed \ct{1A-H} referred to the modern cosmological theory,
which assumes the existence of a lot of degenerate vacua in the Universe.\\
In Ref.\ct{1V} the existence of different vacua of our Universe was explained by the MPP.
It was shown that these vacua are regulated by the baryon charge and all coexisting vacua
exhibits the baryon asymmetry. The present baryon asymmetry of the Universe was discussed.\\
MPM is a base for the new model of the Higgs inflation developed in the present paper.

\section{Plebanski's formulation of General Relativity}

Previously we have constructed a model unifying gravity with some,
e.g. weak SU(2) gauge and the "Higgs" scalar fields - so called
Gravi-Weak Unification (GWU) model \ct{1u}. We developed our GWU
model in accordance with a general model of unification of
gravity, gauge and Higgs fields, suggested in Ref.~ \ct{2u}.
Constructing the GWU-model, we have used an extension of the
Plebanski's 4-dimensional gravitational theory \ct{3P}, in which
the fundamental fields are two-forms containing tetrads and spin
connections, and in addition, certain auxiliary fields.

Theory of General Relativity (GR) was formulated by Einstein as
dynamics of the metrics  $g_{\mu \nu}$. Later, Plebanski \ct{3P},
Ashtekar \ct{4P} and other authors \ct{6P,7P} presented GR in a
self-dual approach, in which the true configuration variable is a
self-dual connection corresponding to the gauging of the local
Lorentz group, $SO(1,3)$, or the spin group, $Spin(1,3)$.

In the Plebanski's formulation of the 4-dimensional theory of
gravity \ct{3P,4P,6P,7P}, the gravitational action is the product
of two 2-forms, which are constructed from the connections
$A^{IJ}$ and tetrads (or frames) $e^I$ considered as independent
dynamical variables. Both $A^{IJ}$ and $e^I$ are 1-forms:
\be   A^{IJ} =  A_{\mu}^{IJ}dx^{\mu} \quad {\mbox{and}}\quad
       e^I = e_{\mu}^Idx^{\mu}.
                          \lb{1} \ee
Also 1-form \be A = \frac 12 A^ {IJ}\gamma_ {IJ}  \lb{2} \ee is
used, in which the generators $\gamma_ {IJ}$ are products of
generators of the Clifford algebra $Cl(1,3)$: \be \gamma_ {IJ} =
\gamma_{I}\gamma_{J},  \lb{3} \ee where $\gamma_{I}$ is the Dirac
matrix.

Indices $I, J=0,1,2,3$ belong to the space-time with
Minkowski's metrics $\eta^ {IJ} = {\rm diag} (1,-1,-1,-1)$, which
is considered as a flat space, tangential to the curved space with
the metrics $g_{\mu\nu}$. In this case connection belongs to the
local Lorentz group  $SO(1,3)$, or to the spin group $Spin(1,3)$.

In general case of unifications of gravity with the $SO (N)$, or
$SU (N)$, gauge and Higgs fields, the gauge algebra is $\mathfrak
g = \mathfrak {spin}(p, q)$, where $p=q=1+N$ and we have $I, J =
1,2,...,p+q$. In our model of unification of gravity with the weak
$SU(2)$ interactions we considered a group of symmetry with the
Lie algebra $\mathfrak {spin}(4,4)$ (the group $Spin(4,4)$ is
isomorphic to $SO(4,4)$-group). In this model, indices $I,J$ run
over all $8\times 8$ values: $I, J = 1,2,..,7,8$.

For the purpose of construction of the action for any unification
theory, the following 2-forms are also considered:
\be
 B^{IJ} = e^I\wedge e^J = \frac 12
e_{\mu}^Ie_{\nu}^Jdx^{\mu}\wedge dx^{\nu},\qquad F^{IJ} = \frac 12
F_{\mu\nu}^{IJ}dx^{\mu}\wedge dx^{\nu}, \lb{4} \ee where \be
F_{\mu\nu}^{IJ} =\partial_{\mu}A_{\nu}^{IJ} -
\partial_{\nu}A_{\mu}^{IJ} + \left[A_{\mu}, A_{\nu}\right]^{IJ}
\lb{5} \ee determines the Riemann-Cartan curvature: \be R_{\kappa
\lambda \mu \nu} = e_{\kappa}^I e_{\lambda}^JF_{\mu\nu}^{IJ}.
\lb{6} \ee Also 2-forms $B$ and $F$ are considered : \be B= \frac
12 B^{IJ}\gamma_{IJ}, \qquad  F= \frac 12 F^{IJ}\gamma_{IJ},
\qquad  F = dA + \frac 12 \left[A, A\right]. \lb{9} \ee
The well-known in literature Plebanski's $BF$-theory is submitted
by the following gravitational action with nonzero cosmological
constant $\Lambda$:
\be  I_{(GR)} = \frac{1}{\kappa^2}\int
\epsilon^{IJKL}\left(B^{IJ}\wedge
    F^{KL} + \frac{\Lambda}{4}B^{IJ}\wedge B^{KL}\right),
                                  \lb{11} \ee
where $\kappa^2=8\pi G_N$, $G_N$ is the Newton's gravitational
constant, and $M_{Pl}^{red.} = 1/{\sqrt{8\pi G_N}}$ is the reduced
Planck mass.

Considering the dual tensors:
\be
F^*_{\mu\nu}\equiv \frac
{1}{2\sqrt{-g}}\epsilon^{\rho\sigma}_{\mu\nu}F_{\rho\sigma}, \quad
A^{\star IJ} = \frac 12 \epsilon^{IJKL}A^{KL}, \lb{12} \ee
we can determine self-dual (+) and anti-self-dual (-) components
of the tensor $A^{IJ}$:
\be A^{(\pm)\,IJ}=\left({\cal P}^{\pm}A\right)^{IJ} = \frac 12
\left(A^{IJ} \pm iA^{\star\,IJ}\right).
                                 \lb{14} \ee
Two projectors on the spaces of the so-called self- and
anti-self-dual tensors $${\cal P}^{\pm}= \frac 12\left(\delta^{IJ}_{KL}
\pm  i\epsilon^{IJ}_{KL}\right)$$ carry out the following homomorphism:
\be
   \bf{so}(1,3) = \bf{sl}(2,C)_R \oplus \bf{sl}(2,C)_L, \lb{15} \ee
where $R,L$ mean Right and Left, respectively.

As a result, Eq.(\ref{15}) gives that the non-zero components of
connections are only:
\be A^{(\pm) i } = \pm 2A^{(\pm) 0i}, \lb{16} \ee
where $I=0,i$, and $i=1,2,3$.

Instead of the (anti-)self-duality, the terms of the left-handed
(+), or $L$, and right-handed (-), or $R$, components are used.

Plebanski \ct{3P} and the authors of Refs. \ct{4P,6P,7P} suggested
to consider a gravitational action in the visible O-world as a
left-handed $\mathfrak{sl}(2,C)_L^{(grav)}$-invariant action,
which contains only self-dual fields $F=F^{(+)i}$ and
$\Sigma=\Sigma^{(+)i}$ (i=1,2,3):
\be I_{(grav)}(\Sigma,A,\psi) = \frac{1}{\kappa^2} \int
\left[\Sigma^i\wedge F^i +
 \left(\Psi^{-1}\right)_{ij}\Sigma^i\wedge \Sigma^j\right]
                      \lb{18} \ee
with
\be (\Psi^{-1})_{ij} = \psi_{ij} - \frac{\Lambda}{6}\delta_{ij}.
 \lb{18a} \ee
Here $\Sigma^i=2B^{0i}$, and $\Psi_{ij}$ are auxiliary fields
defining a gauge, which provides an equivalence of Eq.~(\ref{18})
to the Einstein-Hilbert gravitational action:
\be I_{(EG)} = \frac{1}{\kappa^2} \int d^4 x (\frac{R}{2} -
\Lambda ),
                           \lb{19} \ee
where $R$ is a scalar curvature, and $\Lambda$ is the Einstein
cosmological constant.

It was assumed in Refs. \ct{3P,4P,6P,7P} that the anti-self-dual
right-handed gravitational world is absent in the Nature (
$\Sigma^{(-)} = F^{(-)} = 0$), and our world, in which we live, is
a self-dual left-handed gravitational world described by the
action (\ref{18}).

Following the ideas of Ref.~\ct{3}, we distinguish the two worlds
of the Universe, visible and invisible, and consider the two
sectors of gravity: left-handed gravity and right-handed gravity.

If there exists in the Nature a duplication of worlds with opposite
chiralities - Ordinary and Mirror -- we can consider the
left-handed gravity in the Ordinary world and the right-handed
gravity in the Mirror world.
The anti-self-dual right-handed gravitational action of the mirror
world MW is given by the following integral:
\be I^{(grav)}_{(MW)}(\Sigma^{(-)},A^{(-)},\psi') =
\frac{1}{{\kappa'}^2} \int [\Sigma^{(-)i}\wedge F^{(-)i} +
 ({\Psi'}^{-1})_{ij}\Sigma^{(-)i}\wedge \Sigma^{(-)j}].
                      \lb{21} \ee
In Eqs.~(\ref{18}) and (\ref{21}) we have:
\be  \Sigma^{(\pm) i} = e^0\wedge e^i \pm i\frac 12
     \epsilon^i_{jk}e^j\wedge e^k.
                          \lb{22} \ee
A correct gauge was provided by Plebanski, when he introduced in
the gravitational action the Lagrange multipliers $\psi_{ij}$ - an
auxiliary fields, symmetric and traceless. These auxiliary fields
provide a correct number of constraints.

Plebanski considered the action (\ref{18}) with the following
constraints:
\be  \Sigma^i\wedge \Sigma^j - \frac 13 \delta^{ij}\Sigma^k\wedge
\Sigma_k = 0,   \lb{24} \ee
and
\be  \Sigma^i\wedge \Sigma^{(-)j} = 0.  \lb{25} \ee
The variables $\Sigma_{\mu\nu}^i$ have 18 degrees of freedom. The
five conditions (\ref{24}) leave 13 degrees of freedom, and the
condition (\ref{25}) leaves 10 degrees of freedom, which coincides
with a number of degrees of freedom given by the metric tensor
$g_{\mu\nu}$. This circumstance confirms the equivalence of the
actions (\ref{18}) and (\ref{19}). And now there exist the
following Lie algebra describing the GR:
\be
   \mathfrak{so}(1,3) = \mathfrak{sl}(2,C)_L \oplus
   \mathfrak{sl}(2,C)_R.
                                 \lb{26} \ee
But later, in Ref.~\ct{4}, we assumed that the mixing of
$g_{\mu\nu}^{L,R}$ can be so strong that the left-handed gravity
coincides with the right-handed gravity: the left-handed and
right-handed connections are equal, that is, $A_L=A_R$. In this
case $g_{\mu\nu}^L = g_{\mu\nu}^R$, and the left-handed and
right-handed gravity equally interact with the visible and mirror
matters.

The same approach was considered in the present paper.

\section{Gravi-Weak Unification model}

On a way of unification of the gravitational and weak interactions
we considered in Ref. \ct{1u} an extended $\mathfrak g = \mathfrak
{spin}(4,4)$-invariant Plebanski's action:
\be I(A, B, \Phi) = \frac{1}{g_{uni}} \int_{\bf
M}\left\langle BF +  B\Phi B + \frac 13 B\Phi \Phi \Phi B
\right\rangle, \lb{22u} \ee
where $\langle...\rangle$ means a wedge product, $g_{uni}$ is an
unification parameter, and $\Phi_{IJKL}$ are auxiliary fields.
Here $I,J,K,L = 1,2,...7,8$.

Having considered the equations of motion, obtained by means of
the action (\ref{22u}), and having chosen a possible class of
solutions, we can present the following action for the Gravi-Weak
Unification (GWU) - see details in Refs.~\ct{1u,2u}:
\be   I(A, \Phi) = \frac{1}{8g_{uni}} \int_{\bf M} \langle
\Phi FF \rangle, \lb{38u} \ee
where
\be  \langle \Phi F F \rangle =
\frac{d^4x}{32}\epsilon^{\mu\nu\rho\sigma}{{\Phi_{\mu\nu}}^{\varphi\chi
IJ}}_{KL}F_{\varphi\chi IJ} {F_{\rho\sigma}}^{KL},  \label{39u} \ee
and
\be {{\Phi_{\mu\nu}}^{\rho\sigma ab}}_{cd} = (e_{\mu}^f)
(e_{\nu}^g){\epsilon_{fg}}^{kl}(e^{\rho}_k)
(e^{\sigma}_l)\delta_{cd}^{ab}. \label{40u} \ee
A spontaneous symmetry breaking of our new action that produces
the dynamics of gravity, weak $SU(2)$ gauge and Higgs fields,
leads to the conservation of the following sub-algebra:
$$  \mathfrak {g_1} = {\mathfrak {sl}(2,C)}^{(grav)}_L
\oplus {\mathfrak {su}(2)}_L.$$ Considering indices $a, b \in
\{0,1,2,3\}$ as corresponding to $I,J=1,2,3,4$, and indices $m, n$
as corresponding to indices $I, J=5,6,7,8$, we can present a
spontaneous violation of the Gravi-Weak Unification symmetry in
terms of the 2-forms:
\be A = \frac 12 \omega + \frac 14 E + A_W, \lb{41u} \ee
where $\omega = \omega^{ab}\gamma_{ab}$ is a gravitational
spin-connection, which corresponds to the sub-algebra $\mathfrak
{sl}(2,C)_L^{(grav)}$. The connection $E = E^{am}\gamma_{am}$
corresponds to the non-diagonal components of the $8\times 8$
matrix $A^{IJ}$, described by the following way (see \ct{2u}):
\be
 E = e\varphi = e^a_{\mu}\gamma_a\varphi^m\gamma_m dx^{\mu}. \lb{42u} \ee
The connection $A_W = \frac 12 A^{mn}\gamma_{mn}$ gives: $A_W =
\frac 12 A_W^i \tau_i$, which corresponds to the sub-algebra
$\mathfrak {su}(2)_L$ of the weak interaction ($\tau_i$ are the
Pauli matrices with $i=1,2,3$).

Assuming that we have only scalar field
$\varphi^m=(\varphi,\varphi^i)$, we can consider a symmetry
breakdown of the Gravi-Weak Unification, leading to the following
OW-action \ct{1u}:
$$I_{(OW)}\left(e,\varphi,A,A_W\right)= \frac{3}{8g_{uni}}
\int_{\bf M} d^4x|e|\left(\frac 1{16}{|\varphi|}^2 R -
\frac{3}{32}{|\varphi|}^4 \right.$$
 \be + \frac 1{16}{R_{ab}}^{cd}
{R^{ab}}_{cd} - \left.\frac 12 {\cal D}_a{\varphi^{\dagger}} {\cal
D}^a\varphi - \frac 14 {F_W^i}_{ab}{F_W^i}^{ab} \right). \lb{27u}
\ee
In Eq.~(\ref{27u}) we have the Riemann scalar curvature $R$;
$|\varphi |^2 = {\varphi }^{\dag}\varphi$ is a squared scalar
field, which from the beginning is not the Higgs field of the
Standard Model; ${\cal D }\varphi = d\varphi + [A_W, \varphi]$ is
a covariant derivative of the scalar field, and $F_W = dA_W +
[A_W, A_W]$ is a curvature of the gauge field $A_W$. The third
member of the action (\ref{27u}) is a topological term in the
Gauss-Bone theory of gravity (see for example \ct{3u,3uu}).

Lagrangian in the action (\ref{27u}) leads to the nonzero vacuum
expectation value (VEV) of the scalar field:
$v=\langle\varphi\rangle =\varphi_0$, which corresponds to a local
minimum of the effective potential $V_{eff} (\varphi)$  at $v^2 =
R_0/3$, where $R_0 > 0$ is a constant de Sitter space-time
background curvature \ct{2u}.

According to (\ref{27u}), the Newton gravitational constant $G_N$
is defined by the expression:
\be 8\pi G_N = ({M^{(red.)}_{Pl}})^{-2} = \frac{64g_{uni}}{3v^2},
\lb{28u} \ee
a bare cosmological constant is equal to
\be \Lambda_0 = \frac 34 v^2, \lb{29u} \ee
and
\be g_W^2 = 8g_{uni}/3. \lb{30u} \ee
The coupling constant $g_W$ is a bare coupling constant of the
weak interaction, which also coincides with a value of the
constant $g_2=g_W$ at the Planck scale. Considering the running
$\alpha_2^ {-1}(\mu)$, where $\alpha_2=g_2^2/4\pi$, we can carry
out an extrapolation of this rate to the Planck scale, what leads
to the following estimations \ct{4u}:
\be \alpha_2 (M_{Pl}) \sim 1/50, \, \, g_{uni}\sim 0.1. \lb{31u}
\ee
Having substituted in Eq.(\ref{28u}) the values of $g_ {uni}
\simeq 0.1$ and $G_N=1/8\pi {(M_{Pl}^{red.})}^2$, where $M_{Pl}^{
red.}\approx 2.43\cdot 10^{18}$ GeV, it is easy to obtain the
VEV's value $v$, which in this case is located near the Planck
scale:
\be v=v_2\approx 3.5\cdot 10^{18} {\rm{GeV}}. \lb{32u} \ee
Such a result takes place, if the Universe at the early stage
stayed in the "false vacuum", in which the VEV of the Higgs field
is huge: $v=v_2\sim 10^{18} GeV$. The exit from this state could
be carried out only by means of the existence of the second scalar
field. In the present paper we assume that the second scalar
field, participating into the Inflation, is the mirror Higgs
field, which arises from the interaction between the Higgs fields
of the visible and invisible sectors of the Universe.

\section{Mirror world with broken mirror parity}

In contrast to the article \ct{2u}, in this paper we assume the
existence in the Nature of the invisible (hidden) Mirror World
(MW) parallel to the visible Ordinary World (OW).

Such a hypothesis was suggested in Refs.~\ct{LY,KOP}.

The Mirror World (MW) is a mirror copy of the Ordinary World (OW)
and contains the same particles and types of interactions as our
visible world, but with the opposite chirality. Lee and Yang
\ct{LY} were first to suggest such a duplication of the worlds,
which restores the left-right symmetry of the Nature. The term
``Mirror Matter'' was introduced by Kobzarev, Okun and Pomeranchuk
\ct{KOP}, who first suggested to consider MW as a hidden
(invisible) sector of the Universe, which interacts with the
ordinary (visible) world only via gravity, or another (presumably
scalar) very weak interaction.

In the present paper we consider the hidden sector of the Universe
as a Mirror World (MW) with broken Mirror Parity (MP)
\ct{1mw,2mw,3mw,4mw,5mw}. If the ordinary and mirror worlds are
identical, then O- and M-particles should have the same
cosmological densities. But this is immediately in conflict with
recent astrophysical measurements \ct{1D,2D,3D}. Astrophysical and
cosmological observations have revealed the existence of the Dark
Matter (DM), which constitutes about 25\% of the total energy
density of the Universe. This is five times larger than all the
visible matter, $\Omega_{DM}: \Omega_{M} \simeq 5 : 1$. Mirror
particles have been suggested as candidates for the inferred dark
matter in the Universe \ct{4mw,7mw} (see also \ct{Sil}).
Therefore, the mirror parity (MP) is not conserved, and the OW and
MW are not identical.

In the Refs. \ct{1mw,2mw,3mw,4mw,5mw} it was suggested that the
VEVs of the Higgs doublets $\phi$ and $\widetilde{\phi}$ are not
equal:
\be \langle\phi\rangle=v, \quad \langle\tilde{\phi}
\rangle=\tilde{v}, \quad {\rm {and} }\quad v\neq \tilde{v}.
\lb{1mw} \ee
The parameter characterizing the violation of the MP is
\be \zeta = \tilde{v}/{v} \gg 1. \lb{2mw} \ee
The parameter $\zeta$ was introduced and estimated in Refs.~\ct{10mw,11mw}.
The estimation gave:
\be \zeta > 30, \quad \zeta \sim 100. \lb{3mw} \ee
Then the masses of mirror fermions and massive bosons are scaled
up by the factor $\zeta$ with respect to the masses of their
OW-counterparts:
\be \widetilde{m}_{\widetilde{q},\widetilde{l}} = \zeta m_{q,l},\qquad
  \widetilde{M}_{\widetilde{W},\widetilde{Z},\widetilde{\phi}} =
  \zeta  M_{W,Z,\phi}, \lb{4mw} \ee
while photons and gluons remain massless in both worlds.

\subsection{Communications between visible and hidden worlds}

The dynamics of the two worlds of the Universe, visible and
hidden, is governed by the following action:
\be   I = \int d^4x|e|\left[ L_{(grav)}  + L_{SM} +
\widetilde{L}_{\widetilde{SM}} + L_{(mix)}\right], \lb{5mw} \ee
where $L_{(grav)}$ is the gravitational Lagrangian, $L_{SM}$ and
$\widetilde{L}_{\widetilde{SM}}$ are the SM Lagrangians in the OW
and MW, respectively.

$L_{(mix)}$ is the Lagrangian describing all mixing terms giving
small contributions to physical processes: mirror particles have
not been seen so far, and the communication between visible and
hidden worlds is hard.

In Ref.~\ct{12mw} it was assumed that along with gravitational
interaction between OW and MW worlds, there also exists the
interaction between the initial Higgs field $\phi$ and the mirror
Higgs field $\widetilde{\phi}$:
\be V_{int} = \alpha_h(\phi^{\dagger} \phi)( \widetilde{\phi}
^{\dagger}\widetilde{\phi}), \lb{6mw} \ee
where $\phi$ ($\widetilde{\phi}$) is the SM ($\widetilde{SM}$)
Higgs doublet. This interaction exists in $L_{mix}$, and the
coupling constant $\alpha_h$ is assumed as very small.

\subsection{Gravi-Weak action in the invisible (mirror) sector of the Universe}

The action $I_{(MW)}$ in the mirror world is represented by the
same integral  (\ref{27u}), in which we have to make the
replacement of all OW-fields by their mirror counterparts: $$e,
\phi, A, A_W, R \to \widetilde{e}, \widetilde{\phi},
\widetilde{A}, \widetilde{A}_W, \widetilde{R}.$$ However:
$$\widetilde{g_{uni}} = g_{uni},$$ because we assume that at
the early stage of the evolution of the Universe, mirror parity
was NOT broken.

\section{U(4)-group of fermions}

Constructing the unification of the Gravity and Standard Model
(SM) gauge groups by using algebraic spinors of the standard
four-dimensional Clifford algebra with a left-right symmetry, we
imagine the creation of the SM families at the Planck scale as it
was suggested in the theory \ct{1S} and in the RD \ct{1R,2R}.

We assume that at the early stage of evolution of the Universe
(say, in $\sim 10^{-43}$ sec after the Big Bang) the direct
product of the gauge groups of GWU and internal symmetry $U(4)$
randomly emerges:
\be G_{(GW)} \times U(4), \lb{34f} \ee
which further was broken due to the following breaking chains:
\be G_{(GW)} \to SL (2,C)^{(grav)} \times SU(2)^{(weak)}, \lb{35f}
\ee
and
\be U(4) \to SU(4) \times U(1)_{(B-L)} \to SU(3) \times U(1)_Y
\times U(1)_{(B-L)}. \lb{36f} \ee
Below the see-saw scale ($M_R\sim 10^{12}$ GeV) we obtain the following group of symmetry:
\be
SL(2,C)^{(grav)} \times SU(3)_{(color)} \times SU(2)^{(weak)}
            \times U(1)_Y, \lb{37f} \ee
or \be SL (2,C)^{(grav)} \times G_{(SM)}. \lb{38f} \ee
Spinors appear in multiplets of gauge groups.

We can consider the 1-form connection, $A =\frac 12
A^{IJ}\gamma_{IJ}$, as an independent physical variable describing
the geometry of the d-dimensional space-time. The  curvature   $$F
= dA + \frac 12 [A, A]$$ and $$B=\frac 12 B^{IJ}\gamma_{IJ}$$ are
$\mathfrak{spin}(p,q)$-valued 2-form fields. The generators
$\gamma_{IJ}=\gamma_{I}\gamma_{J}$ of the
$\mathfrak{spin}(p,q)$-algebra have indices running over all
$(p+q)\times (p+q)$ values: $I,J = 1,2,...,p+q$.

In Ref.~\ct{1u}, in correspondence with  Ref.~\ct{2u}, we have
developed the Gravi-Weak Unification model starting with the
$\mathfrak g = \mathfrak {spin}(4,4)$-invariant extended
Plebanski's action (\ref{27u}).

The standard four-dimensional Clifford algebra $Cl_{1,3}$ is given
by Dirac gamma matrices, $\{\gamma_{\mu},\gamma_{\nu}\}=2
\eta_{\mu\nu}$, in Weyl representation, with $\eta_{\mu\nu}={\rm
diag}(-1,1,1,1)$ and $\gamma_5={\rm diag}(1,1, -1, -1)$. For
$Cl_{1,3}$ we use the basis $$\{\gamma_{A=1,..,16}\}=
\{1_4,\,\gamma_0,\,\gamma_i,\,\gamma_0\gamma_i,\,
-\gamma_i\gamma_j,\,\gamma_5\gamma_0,\,\gamma_5\gamma_i,\,\gamma_5\},$$
with i=1,2,3. This is also the algebra $\mathfrak {gl}(4,C)$ of
all $4\times 4$ matrices.

The aim to set quantum numbers of fermions in families of the SM
gave the traditional way to embed the SM family into
Grand-Unification groups (like $SU(5), SO(10), E_6$, etc.) All
these approaches consider the gauge groups as internal symmetries.
Fermions appear in multiplets of the internal groups, forming a
direct product with the space-time Lorentz group.

However, there exists a different way: to resort to the algebraic
spinor theory  (see for example Ref.~\ct{2gw}), and to set
fermions in multiplets of the Clifford algebra, which is
isomorphic to the algebra of inhomogeneous differential forms.
These spinors still satisfy the Dirac equation. They are not in
the minimal representation of the Lorentz group: they are generic
elements of the Clifford algebra -- objects of dimension $2^d$.
These multiplets naturally can contain more particles, than the
usual description gives, and can accommodate various sets of
quantum numbers, including the SM ones. As a result, this approach
leads to a promising way to unify gravity with gauge interactions.

Usual spinors $\psi$ are column objects transforming under Lorentz
transformations, while algebraic spinors are objects of the
Clifford algebra: $\Psi=\psi^A\gamma_A$, which are represented by
$4\times 4$ complex matrices $\Psi=\{\psi_{ab}\}$, with
a,b=1,2,3,4.

These objects are transformed at the left under the algebra-valued
transformations. Such transformations act on each column
separately, therefore the four columns inside the algebraic spinor
represent four invariant subspaces. Thus, an algebraic spinor
contains four Dirac spinors. These objects belong to another
$\mathfrak{\widetilde {gl}}(4,C)$-algebra that can accommodate an
internal symmetry up to a rank 4, for example, $U(4)$.

Considering the left and right chirality objects, we introduce the
left and right complex algebraic spinors $\Psi_{L,R}$. They again
are represented by $4\times 4$ complex matrices. Summarizing,
$\Psi_{L,R}$ belongs to $U(4)_{L,R}$ group, and $\Psi_L$ contains
four isospin doublets of Weyl spinors, which we can identify with
the left-handed SM family.  It is very suggestive to represent
$\Psi_L$ with lepton and colored quark indices:
\be \label{eq:psidef} \Psi_L=\left(
\begin{array}{llll}
 \nu _{\text{L1}} & u_{\text{L1},r} &
   u_{\text{L1},g} & u_{\text{L1},b}
   \\
 \nu _{\text{L2}} & u_{\text{L2},r} &
   u_{\text{L2},g} & u_{\text{L2},b}
   \\
 e_{\text{L1}} & d_{\text{L1},r} &
   d_{\text{L1},g} & d_{\text{L1},b}
   \\
 e_{\text{L2}} & d_{\text{L2},r} &
   d_{\text{L2},g} & d_{\text{L2},b}
\end{array}
\right).   \ee
In the present paper the SM families of ordinary and mirror worlds
(OW and MW) are described in just by a left-right symmetric couple
of such spinors: $\Psi_L$ and $\Psi_R$.

Considering the left chirality in $OW$, we need a breaking of the
$\mathfrak {gl}(4,C)_L$ to ${\mathfrak {g_1}} = {\mathfrak
{sl}(2,C)}^{(grav)}_L \oplus {\mathfrak {su}}(2)_L.$

The transformations belonging to ${\mathfrak {gl}}(4,C)_L$ should
be restricted to be compact, because non-compact internal
symmetries always contain ghosts. This minimal requirement leads
from ${\mathfrak {gl}}(4,C)$ to its maximal compact group $U(4)$,
i.e. a group that unifies color and $(B - L)$-quantum number by
treating lepton number as the fourth color. The representation
(\ref{eq:psidef}) of $\Psi_L$ explicitly shows this.

In the broken phase of the GWU, the symmetry in the OW would thus
be:
$$ SU(2)^{(grav)}_L \times SU(2)_L \times U(4)_L.$$
Then, the right chirality of the MW would give rise to a second
copy of these:
$$ SU(2)^{(grav)}_R \times SU(2)_R \times U(4)_R.$$
And we have a duplication of worlds.

These groups can be linked to the SM groups by standard breaking
chains:
$$ U(4)\to SU(4)\times U(1)_Y \to SU(3)_c \times
U(1)_{(B-L)}\times U(1)_Y, $$ after introducing appropriate Higgs
fields needed for the symmetry breaking.

Now it is useful to show explicitly the generators of ${{\mathfrak
sl}(2,C)}^{(grav)}$ and ${\mathfrak su}(2)^{(weak)}$:
$$ {{\mathfrak sl}(2,C)}^{(grav)}: \quad \{\rho_i\}=\{\sigma_i\otimes
1_2\},$$ and \be {{\mathfrak su}(2)}^{(weak)}:\quad
\{\tau_i\}=\{1_2\otimes \sigma_i\}.    \lb{22f} \ee
The gauging of symmetries is realized by introducing a covariant
derivative with Clifford-algebra valued vector fields:
\be \partial_{\mu} \to D_{\mu}^{L.R} =  \partial_{\mu} +
    V_{\mu}^{L,R}  + \tilde V_{\mu}^{L,R}, \lb{23f} \ee
where $\tilde V_{\mu}^{L,R}$ are just the gauge fields in
$\mathfrak {u(4)}_{L,R}$-Clifford algebra notation, and the fields
$V_{\mu}^{L,R}$ unify gravity and weak-isospin.

The $V_{\mu}^{L,R}$  can be parametrized in terms of the complex
gauge fields $\omega^L, \bar \omega^R, W^{L,R}$:
$$  V_{\mu}^L = i\omega_{\mu}^i\rho_i + iW_{\mu}^{Li}\tau_i, $$
\be  V_{\mu}^R = i\bar \omega_{\mu}^i\rho_i + iW_{\mu}^{Ri}\tau_i,
\lb{24f}   \ee
where $\omega_{\mu}^i,\,\bar \omega_{\mu}^i$ reproduce the
self-dual and anti-self-dual spin connections of gravity.

Now we can  build kinetic terms for left and right fermions (tr is
the trace in $4\times 4$ representation):
\be L_{kin}= tr[\Psi_L^{\dag}\partial^{\mu} \Gamma_{\mu}^L\Psi_L]
+ tr[\Psi_R^{\dag}\partial^{\mu} \Gamma_{\mu}^R\Psi_R], \lb{25f}
\ee
where:
\be \Gamma_{\mu}^{L,R}=\{\pm 1_2,\sigma_i\}\otimes 1_2. \lb{26f}
\ee

\section{Multiple Point Model and the prediction of the top and Higgs Masses}

Recently discovered Higgs boson \ct{1Hig,2Hig} showed an
intriguing property: among the many different values of the Higgs
mass, which were available, the Nature has chosen one that allows
us to think that the SM is valid up to the Planck scale, apart
from the existence of some right-handed heavy Majorana neutrinos
at a see-saw scale $(\sim 10^{12} - 10^{14}\,\, {\rm{GeV}})$. Of
course, this scenario suffers from the hierarchy problem of why
the ratio of the Planck scale to the Electroweak (EW) scale is so
huge (see \ct{9}). The mechanism for fine-tuning of coupling
constants of the SM was suggested in Refs.~\ct{9,1nbs,2nbs,3nbs}
which give an explanation, why the ratio of these two scales
should be of the order of $10^{17}$, as observed. This is based on
the so-called Multiple Point Principle \ct{1M}.

Some time ago, in Ref.~\ct{2M}, the MPP was applied to the SM by
the consideration of the two degenerate vacua: a vacuum at the
Planck scale with a Higgs field vacuum expectation value (VEV):
$v_2 \sim 10^{18}$ GeV, and EW vacuum having VEV: $v_1 = 246$ GeV.
Consequently the top quark and Higgs boson couplings were
fine-tuned to lie at a point on the SM vacuum stability curve
\ct{26Fr,27Fr,28Fr,29Fr,30Fr,31Fr}. This gave the following MPP
prediction by C.D.~Froggatt and H.B.~Nilsen (see \ct{2M}) for the
top quark and the Higgs boson masses:
\be m_t = 173 \pm 5 GeV, \quad m_H = 135 \pm 9 GeV. \lb{1x} \ee
Later, the prediction for the mass of the Higgs boson was improved
by the calculation of the two-loop radiative corrections to the
effective Higgs potential \cite{9,32Fr,33Fr,34Fr,35Fr}. The
predictions: 125 GeV $\lesssim m_H \lesssim 143$ GeV in
Ref.~\ct{9}, and 129 $\pm \, 2$ GeV in Ref.~\ct{34Fr} -- provided
the possibility of the theoretical explanation of the value
$M_H\approx 126$ GeV observed at the LHC. The authors of the
recent paper \ct{35Fr} have shown that the most interesting aspect
of the measured value of $M_H$ is its near-criticality. They have
thoroughly studied the condition of near-criticality in terms of
the SM parameters at the high (Planck) scale. They extrapolated
the SM parameters up to large energies with full 3-loop NNLO RGE
precision. All these results mean that the radiative corrections
to the Higgs effective potential lead to the true value of the
Higgs mass existing in the
Nature.\\
Using the top quark pole mass $m_t = 173.1 \pm 0.7$ GeV as input
leads (see \ct{34Fr}) to an updated MPP prediction for the Higgs
mass of $m_H = 129.4 \pm 1.8$ GeV. This is very close to the
observed Higgs mass \ct{1Hig,2Hig}: $M_H \approx 126$ GeV. The
result is rather sensitive to the top quark pole mass: a change of
$\Delta m_t = \pm 1$ GeV gives a change in the predicted Higgs mass of
$\Delta m_H = \pm 2$ GeV.\\
We see that in the assumption of the validity of the SM at very
high Planck-mass scales, the measured value of the Higgs mass lies
on, or is very close to the vacuum stability curve. Of course,
vacuum stability curve could be a pure coincidence. However, we
take the attitude that it is not accidental and requires an explanation.
This implies two points:\\
1. The SM should not be modified so much by new physics on the energy range
between the EW scale and the Planck scale, where the second vacuum exists,
that the  renormalization group running of the Higgs quartic coupling
$\lambda (\mu)$ is significantly altered.\\
2. There must for some reason exist in the Nature a physical
principle forcing one vacuum to be so closely degenerate with
another one that it is barely stable, or just metastable \ct{FNT}.
Such a principle is, of course, the above-mentioned MPP. MPP is
really a mechanism for fine-tuning couplings.

In order to fine-tune the SM couplings so as to generate the large
ratio of the Planck scale to the EW-scale using MPP, it is
necessary for them to produce a third vacuum in the SM with zero
energy density, i.e. to consider the triple point in the phase
diagram of our theory \ct{9,1nbs,2nbs,3nbs}. In such a speculative
picture this new vacuum is formed at the EW scale by a Bose
condensation of a strongly bound state of 6 top and 6 anti-top
quarks \ct{2nbs,9,4nbs,5nbs,6nbs}.

The behavior of the Higgs self-coupling $\lambda$ is quite peculiar:
it decreases with energy to eventually arrive to a minimum at the
Planck scale values and then starts to increase there after. Within
the experimental and theoretical uncertainties the Higgs coupling
$\lambda$ may stay positive all way up till the Planck scale, but it may
also cross zero at some scale $\mu_0$. If that happens, our Universe
becomes unstable.\\
The largest uncertainty in couplings comes from the determination
of the top Yukawa coupling. Smaller uncertainties are associated
to the determination of the Higgs boson mass and the QCD coupling
$\alpha_s$ (see Refs.~\ct{9,32Fr,33Fr,34Fr,35Fr}).\\
Calculations of the lifetime of the SM vacuum are extremely
sensitive to the Planck scale physics. The authors of
Refs.\ct{1Br,2Br,3Br} showed that if the SM is valid up to the
Planck scale, then the Higgs potential becomes unstable at $\sim
10^{11}$ GeV. There are two reasons of this instability. In
typical tunnelling calculations, the value of the field at the
center of the critical bubble is much larger than the point of the
instability. In the SM case, this turns out to be numerically
within an order of magnitude of the Planck scale.

The measurements of the Higgs mass and top Yukawa coupling
indicate that we live in a very special Universe: at the edge of
the absolute stability of the EW vacuum. If fully stable, the SM
can be extended all the way up to the inflationary scale and the
Higgs field, non-minimally coupled to gravity with strength $\xi$,
can be responsible for the inflation (see
Refs.~\ct{1Bez,2Bez,3Bez}).

\section{The Higgs Inflation model from the "false vacuum" and the mirror
Higgs boson}

The most interesting property of the Universe is the relation
between the particle physics and cosmology: between the elementary
particles theory and the structure of the Universe.\\
The compatibility of the modern astrophysical data with vacuum
stability (or instability) is one of the most important problem
for invoking new physics beyond the SM. In particular, it was
suggested in Refs.\cite{1Bez,2Bez,3Bez} that the Higgs inflation
scenario, in which the Higgs field is non-minimally coupled to
gravity with strength $\xi$, cannot take place if the Higgs
self-coupling $\lambda(\mu)$ (here $\mu$ is an energy scale)
becomes negative at some $\mu_0$
below the inflationary scale.\\
In our model, presented here, we assume that at the early stage of
the evolution of the Universe, the discrete space-time randomly
emerged the group of symmetry $G = G_{(GW)} \times U(4)$ (via the
SUSY GUT group or not), where $G_{(GW)}$ is the Gravi-Weak
Unification group considered in Section 4. In subsequent evolution
of the Universe the temperature decreases and the group $G_{(GW)}$
undergoes the breakdown to the symmetry given by Eq.(\ref{35f}).
As a result, the obtained Lagrangian (exactly the effective
potential $V_{eff} (\varphi)$) shows that the vacuum with the VEV
$v=v_2$ of the scalar field $\varphi$ (so called "second vacuum")
appears at large field values $\sim 10^{18}$ GeV. Following to the
MPM scenario (see Section 7), the Higgs effective potential has
the next minimum at the EW scale, and the vacuum with the VEV
equal to $v_1 = 246$ GeV
(so called "first vacuum") corresponds to the vacuum, in which we live now.\\
The Higgs inflation scenario is heavily based on the standard
vacuum stability analysis. In particular, it requires that a new
physics shows up only at the Planck scale $M_{Pl}$, and that the
SM lives at the edge of the stability region, where
\be \lambda (M_{Pl}) \sim 0 \quad \mbox{and} \quad \beta (\lambda
(M_{Pl})) \sim 0. \lb{45} \ee
Here $\beta (\lambda)$ is the beta-function of the renormalization
group equation (RGE) for the Higgs self-coupling constant
$\lambda$. We see, however, that the new physics interactions at
the Planck scale can strongly change this situation.\\
The realization of the conditions (\ref{45}) requires such a fine tuning that
even a small grain of new physics at the Planck scale can totally destroy the picture.
In our model such a fine tuning is MPP, considered in Section 7.\\
We believe that our model makes a chance for the realization of the Higgs inflation
scenario, showing the relation between low and high energy parameters.\\
The possibility that the SM is valid up to the Planck scale,
$M_{Pl}$, is nowadays largely explored. For example, many papers
were devoted to the scenarios of the Higgs inflation, using the
above mentioned assumptions (see for example
Refs.~\ct{1Br,2Br,3Br,1Bez,2Bez,3Bez,1MT,1H,2H}, etc.).\\
In the present paper we develop an alternative possibility (different with all
previous considerations), using an additional mirror scalar field $\tilde{\varphi}$,
which is very weakly coupled with the usual Higgs field.\\
We start with a local minimum of the Higgs potential at the field
value VEV $v=v_2 \sim 10^{18}$ GeV, as it was predicted by our GWU
model (see Section 4). This minimum exists for a narrow band of
the top quark and Higgs mass values. Not only such a local minimum
exists, but within the allowed parameter range in the top-Higgs
masses, this minimum has the right value of the energy density,
which gives rise to the correct amplitude of density
perturbations.

The existence of the "false vacuum" ($v=v_2\sim 10^{18}$ GeV) in
the Higgs potential, which is a source of the exponential
expansion in the early Universe, gives not only a combined
prediction of the top-Higgs masses, but also the ratio
of tensor to scalar perturbations (rate $r$) from the Inflation.\\
The possibility of achieving a transition from the exponential expansion to
a hot radiation era is provided by adding an extra scalar in the gravitational
sector of the theory, which slows down the expansion of the Universe.

The height of the potential at the time at which density
perturbations are produced is given by the SM Higgs field. The
combined predictions on $m_H,\,m_t$ and $r$ are generic and
independent on the way, what exit from Inflation is realized.

In fact, we present here an alternative model, which provides a
graceful exit from the Inflation and gives rise to a radiation
era.

In order to achieve enough Inflation, we have the scalar field
$\varphi$ trapped at the value $\varphi_0$ (of order $\sim
10^{18}$ Gev) with a suppressed tunnelling rate ($\Gamma <<
H_0^4$, where $\Gamma$ is the tunnelling probability per unit time
and value, and $H_0$ is the Hubble rate).

Then it is necessary, after at least 50-60 e-folds, to trigger a
phase transition through an additional scalar field. In our
GWU-model, this is a mirror scalar field $\tilde{\varphi}$, which
plays a role of a clock, so that $\varphi$ is not stuck at
$\varphi_0$ anymore, and exits through the classical rolls (not
through tunnelling, which is extremely small).\\
When this happens, the field $\varphi$ can roll down fast to smaller values, ending
thus Inflation.\\
Here it is necessary to emphasize, that in our GWU-model the field
$\varphi$ is not a Higgs doublet field of $SU(2)_L$, as in the SM,
but it is an adjoint field of the $SU(2)_L$-group of the weak
interactions as in Refs.~\ct{2u,1Lisi}. We assume, that this field
$\varphi$ decays during the Inflation into the two SM Higgs
doublet fields $\phi$ very quickly, and as a result, the Higgs
field $\phi$ continues the Inflation. At the end, this field
eventually oscillates around zero and dissipate energy, producing
thus a hot plasma, and finally relax at its true minimum at the
usual value $\phi_1=246$ GeV, corresponding to the vacuum, in
which we live now.

Then we choose the following behavior: $\Gamma$ is time-depended,
but $H_0$ stays roughly constant. The Higgs field $\phi$ interacts
directly with an additional mirror Higgs doublet $\widetilde
{\phi}$, as it was suggested in Ref.~\ct{12mw}: the
Foot-Kobakhidze-Volkas formula is (\ref{6mw}). The authors of
Ref.~\ct{12mw} assumed the SM Higgs doublets.

The field $\widetilde {\phi}$ has a time of evolution and modifies
the shape of the barrier in the potential $V_{eff} (\phi)$: the
bottom rises up and then disappears giving the chance to the
inflaton ($\phi$) to roll down toward the first (EW) vacuum.

This so-called Hybrid Inflation scenario was first suggested by A.
Linde in his paper Ref.~\ct{Lin}.

We see, that "waterfall" field $\phi$ is trapped at the almost
zero value and can start evolve, when another field $\tilde \phi$
reaches the almost zero value $\simeq H_0$. Inflation stops, when
$\tilde \phi$ reaches some value $\tilde \phi_{(end)} = \tilde \phi_E$.\\
Such a picture introduces an explicit coupling between the Higgs
field $\phi$ and extra (mirror) field $\widetilde {\phi}$. This is
an explicit coupling constant $\alpha_h$ in Eq.~(\ref{6mw}).

Of course, this picture changes the RGE themselves. However, the
coupling constant $\alpha_h$ needs to be very weak, and the mass
$\widetilde {\phi}$ very large, - then the contribution to the RGE
is very small (practically zero), leaving the connection between
low-energy parameters and the "false vacuum" values unchanged.

\subsection{The Planck scale GWU action in the ordinary world}

In Section 4 we obtained the GWU action given by Eq.~(\ref{27u}).
The gravitational part of the action is:
$$   I_{(OW)}\left(e,\varphi,A,A_W\right)= \frac{3}{8g_{uni}}
\int_{\bf M} d^4x|e|\left(\frac 1{16}{|\varphi|}^2 R -
\frac{3}{32}{|\varphi|}^4
 +...\right) $$\be = \frac{3}{64g_{uni}}
\int_{\bf M} d^4x|e|\left(\frac 12{|\varphi|}^2 R -
\frac{3}{4}{|\varphi|}^4
 +...\right).  \lb{1z} \ee
Considering the background value $R\simeq R_0$, we can find a
minimum of the potential:
\be    V_{eff}(\varphi) \sim - \frac 12{|\varphi|}^2 R_0 +
\frac{3}{4}{|\varphi|}^4
 \lb{2z} \ee
at $\varphi_0=<\varphi>=v$. Here $v^2=R_0/3$. Then according to
(\ref{28u}), we obtain:
\be   I_{(OW)}\left(e,\varphi,A,A_W\right)= \int_{\bf M}
d^4x\sqrt{-g}\left(\frac{M^{red}_{Pl}}{v}\right)^2\left(\frac
12{|\varphi|}^2R - \frac{3}{4}{|\varphi|}^4
 +...\right).  \lb{3z} \ee
In the action (\ref{3z}) the Lagrangian includes the non-minimal
coupling with gravity \ct{1Bez,2Bez,3Bez}.

We see that the field $\varphi$ is not stuck at $\varphi_0$
anymore, but it can be represented as
\be \varphi = \varphi_0 - \sigma = v - \sigma ,  \lb{4z} \ee
where the scalar field $\sigma$ is an {\un {inflaton}}. Here we
see that in the minimum, when $\varphi=v$, the inflaton field is
zero ($\sigma=0$) and then it increases with falling of the field
$\varphi$.

Considering the expansion of the Lagrangian around the background
value $R\simeq R_0$ in powers of the small value $\sigma/v$, and
leaving only the first-power terms, we can present the
gravitational part of the action as:
\be I_{(grav\,\, OW)} = \int_{M}d^4x
\sqrt{-g}\left(M_{Pl}^{red}\right)^2 \left( \frac 12 R_0 |1+
\sigma/v|^2 - \Lambda_0|1 + \sigma/v|^4 + ...\right). \lb{5z} \ee
Here $\Lambda_0 = \frac 34 v^2 = R_0/4$. Using the last relations,
we obtain:
\be I_{(grav\,\, OW)} = \int_{M}d^4x
\sqrt{-g}\left(M_{Pl}^{red}\right)^2 \left( \Lambda_0 -
\frac{m^2}{2}|\sigma|^2 + ...\right), \lb{6z} \ee
where $m^2 = 6$ is the bare mass of the inflaton in units
$M_{Pl}^{red.}=1$.

In the Einstein-Hilbert action the vacuum energy is:
\be \rho_{vac} = \left(M_{Pl}^{red}\right)^2 \Lambda. \lb{7z} \ee
In our case (\ref{6z}) the vacuum energy density is negative:
 \be \rho_0 =
- \left(M_{Pl}^{red}\right)^2 \Lambda_0. \lb{8z} \ee
However, assuming the existence of the discrete space-time of the
Universe at the Planck scale and using the prediction of the
non-commutativity suggested  by B.G. Sidharth \ct{1S,2S}, we
obtain that the gravitational part of the GWU action has the
vacuum energy density equal to zero or almost zero.

Indeed, the total cosmological constant and the total vacuum
density of the Universe contain also the vacuum fluctuations of
fermions and other SM boson fields:
\be \Lambda \equiv \Lambda_{eff} = \Lambda^{ZMD} - \Lambda_0 -
\Lambda^{(NC)}_s + \Lambda^{(NC)}_f, \lb{9z} \ee
where $\Lambda^{ZMD}$ is zero modes degrees of freedom of all
fields existing in the Universe, and $\Lambda^{(NC)}_{s,f}$ are
boson and fermion contributions of non-commutativity. If according
to the theory by B.G. Sidharth \ct{1S}, we have:
\be \rho_{vac}^{(0)} = \left(M_{Pl}^{red}\right)^2 \Lambda^{(0)} =
 \left(M_{Pl}^{red}\right)^2 (\Lambda^{ZMD} -
\Lambda_0 - \Lambda_s^{(NC)}) \approx 0, \lb{10z} \ee
then Eq.~(\ref{6z}) contains the cosmological constant
$\Lambda^{(0)} \approx 0$. In Eqs.~(\ref{9z}) and (\ref{10z}) the
bosonic (scalar) contribution of the non-commutativity is:
\be \rho_{(scalar)}^{(NC)} = m_s^4 \quad ({\rm{in \, units}}:\,
\hbar = c = 1), \lb{11z} \ee
which is given by the mass $m_s$ of the primordial scalar field
$\varphi$. Then the discrete spacetime at the very small distances
is a lattice (or has a lattice-like structure) with a parameter
$$a = \lambda_s = \frac{1}{m_s}.$$ This is a scalar length:
$$
a = \lambda_s \sim 10^{-19}\,\,{\rm GeV}^{-1},$$ which coincides
with the Planck length:
$$
\lambda_{Pl} = \frac{1}{M_{Pl}} \approx 10^{-19}\,\,{\rm
GeV}^{-1}.$$
The assumption:
\be \Lambda^{(0)} = \Lambda^{ZMD} - \Lambda_0 - \Lambda_s^{(NC)}
\approx 0 \lb{12z} \ee
means that the Gravi-Weak Unification model contains the
cosmological constant equal to zero or almost zero.

B.G.~Sidharth gave in Ref.~\ct{11S} the estimation:
\be \rho_{DE} = \left(M_{Pl}^{red}\right)^2 \Lambda_f^{(NC)},
\lb{13z} \ee
considering the non-commutative contribution of light primordial
neutrinos as a dominant contribution to $\rho_{DE}$, which
coincides with astrophysical measurements \ct{1D,2D,3D}:
\be \rho_{DE} \approx (2.3\times 10^{-3}\,\,\rm{eV})^4. \lb{14z}
\ee
Returning to the Inflation model, we rewrite the action (\ref{6z})
as:
\be I_{(grav\,\, OW)} = \int_{\large M} d^4x \sqrt{-g}
\left(M_{Pl}^{red}\right)^2 \left( - \Lambda - \frac{m^2}{2}
|\sigma|^2 + ...\right), \lb{15z} \ee
where the  positive cosmological constant is:
\be \Lambda = \Lambda^{(0)} + \Lambda_f^{(NC)}, \lb{16z} \ee
which is not zero, but very small. The tiny $\Lambda$,
corresponding to Eq.~(\ref{14z}) given by astrophysical
measurements, is a problem of the forthcoming investigations.

We considered the gravitational action in the ordinary world OW.
However, it is quite possible that the mirror world MW exists in
the Nature together with the ordinary world.

\subsection{The Planck scale GWU action of the Universe with
ordinary and mirror worlds}

In this Subsection we present the GWU gravitational action for
both worlds (OW and MW) near their local minima at the Planck
scale.

Taking into account the interaction of the ordinary and mirror
scalar bosons $\varphi$ and $\widetilde{\varphi}$, given by
equation analogous to Eq.~(\ref{6mw}) \ct{12mw}:
\be V_{int} = \alpha_h( \varphi^{\dagger} \varphi)
({\widetilde{\varphi}}^{\dagger}\widetilde{\varphi}), \lb{17z} \ee
we obtain:
$$I_{(grav} = \int_{\large M} d^4x \sqrt{-g}\left[
\left(\frac{M_{Pl}^{red}}{v}\right)^2 \left (\frac 12{|\varphi|}^2
R - \frac{3}{4}{|\varphi|}^4
-\alpha_h{|\varphi|}^2{|\widetilde{\varphi}|}^2
 +...\right)+ ... \right ]$$
 \be +  \int_{\large M} d^4x \sqrt{-g}\left[\left(\frac{{\widetilde M}_{Pl}^{red}}
 {\tilde v}\right)^2 \left (\frac 12{|\widetilde{\varphi}|}^2
{\tilde R} - \frac{3}{4}{|\widetilde{\varphi}|}^4
-\alpha_h{|\varphi|}^2{|\widetilde{\varphi}|}^2
 +...\right) + ... \right].   \lb{18z} \ee
Considering the Planck scale Higgs potential, corresponding to the
action (\ref{18z}), we have:
$$ V(\varphi, \widetilde{\varphi}) \simeq
\left(\frac{M_{Pl}^{red}}{v}\right)^2 \left ( - \frac
12{|\varphi|}^2 R_0  + \frac{3}{4}{|\varphi|}^4
 + \alpha_h{|\varphi|}^2{|\widetilde{\varphi}|}^2 \right)$$ \be +
\left(\frac{{\widetilde M}_{Pl}^{red}}
 {\tilde v}\right)^2 \left( - \frac 12{|\widetilde{\varphi}|}^2
\tilde R_0 + \frac{3}{4}{|\widetilde{\varphi}|}^4 +
{\alpha}_h{|\varphi|}^2{|\widetilde{\varphi}|}^2 \right). \lb{19z}
\ee
According to (\ref{28u}), we have:
\be  \left(\frac{M_{Pl}^{red}}{v}\right)^2 =
\left(\frac{{\widetilde M}_{Pl}^{red}}{\tilde v}\right)^2,
\lb{20z} \ee
and the local minima at $\varphi_0=v$ and ${\widetilde{\varphi}}_0
= \tilde v$ are given by the following conditions:
\be \frac{\partial V(\varphi,
\widetilde{\varphi})}{\partial{|\varphi|^2}}\big|_{|\varphi|=v} =
\left(\frac{M_{Pl}^{red}}{v}\right)^2 \left( - \frac 12 R_0  +
\frac{3}{2}v^2
 + 2\alpha_h{|\widetilde{\varphi}|}^2 \right) = 0,
    \lb{22z} \ee
 \be \frac{\partial V(\varphi,
\widetilde{\varphi})}{\partial{|\widetilde{\varphi}|^2}}\big|_
{|\widetilde{\varphi}|=\tilde
v} = \left(\frac{M_{Pl}^{red}}{v}\right)^2 \left( - \frac 12
\tilde R_0 + \frac{3}{2}{\tilde v}^2 + 2\alpha_h{|\varphi|}^2
\right) = 0,
    \lb{23z} \ee
which give the following solutions:
 \be  v^2 \simeq \frac{R_0}{3} - \frac 43
 \alpha_h{|\widetilde{\varphi}|}^2,
                            \lb{24z} \ee
 \be \tilde v^2 \simeq \frac{\tilde R_0}{3} - \frac
 43\alpha_h{|\varphi|}^2,
                        \lb{25z} \ee
and
 \be  V(\varphi = v, \widetilde{\varphi} = \tilde v) = - \frac 14
\big[ (M_{Pl}^{red})^2R_0 + ({\widetilde M}_{Pl}^{red})^2\tilde
 R_0 \big] = - (M_{Pl}^{red})^2\Lambda_0 -
 ({\widetilde M}_{Pl}^{red})^2\widetilde{\Lambda}_0.
                            \lb{26z} \ee
According to (\ref{28u}), we have:
$$ {\widetilde M}_{Pl}^{red}= \zeta M_{Pl}^{red} \quad
{\rm{and}}\quad \widetilde{\Lambda}_0 = \zeta^2 {\Lambda}_0,
$$
and finally we obtain:
 \be  V(\varphi = v, \widetilde{\varphi} = \tilde v) =
= - (1 + \zeta^4)(M_{Pl}^{red})^2\Lambda_0,
                            \lb{27z} \ee
what gives the negative vacuum energy density. However, as we have
discussed in Subsection 8.1, the cosmological constant is not
given by Eq.~(\ref{27z}). It must be replaced by the cosmological
constant $\Lambda$, which is related with the potential
(\ref{27z}) and the Dark Energy density (\ref{14z}) by the
following way:
 \be  V(\varphi = v,\,\, \widetilde{\varphi} = \tilde v) =
(M_{Pl}^{red})^2\Lambda = \rho_{DE}.
                    \lb{28z} \ee
Using the notation:
 \be \varphi = v - \sigma \quad {\rm{and}}\quad
\widetilde{\varphi} = \tilde v - \widetilde{\sigma}, \lb{29z} \ee
and neglecting the terms containing $\alpha_h$ as very small, it
is not difficult to see that the potential near the Planck scale
is:
 \be  V(\varphi, \,\,\widetilde{\varphi}) =
(M_{Pl}^{red})^2(\Lambda + \frac{m^2}2 |\sigma|^2 + \frac{\tilde
m^2}{2}|\widetilde{\sigma}|^2 + ...),  \lb{30z} \ee
where $m^2 \simeq 6$ and $\tilde m^2 \simeq 6\zeta^2$ (compare
with (\ref{14z})).

The local minimum of the potential (\ref{30z}) at $\varphi_0=v$,
when $\sigma=0$, and ${\widetilde{\varphi}}_0 \neq \tilde v$
($\tilde \sigma \neq 0$) gives:
 \be  V( v,\,\, \widetilde{\varphi})= (M_{Pl}^{red})^2(\Lambda + \frac{\tilde
m^2}2|\widetilde{\sigma}|^2 + ...).
                           \lb{31z} \ee
The last equation (\ref{31z}) shows that the potential $V(v)$
grows with growth of $\tilde \sigma$, i.e. with falling of the
field $\tilde \varphi$. It means that a barrier of potential grows
and at some value $\tilde \sigma = \tilde \sigma|_{in}$ potential
begins its inflationary falling. Here it is necessary to comment
that the position of the minimum also is displaced towards smaller
$\varphi$ (bigger $\sigma$), according to the formula (\ref{24z}).

Our next step is an assumption that during the inflation $\sigma$
decays into the two Higgs doublets of the SM:
\be
      \sigma \to \phi^{\dagger} + \phi.   \lb{32z} \ee
As a result, we have:
 \be \sigma = a_d|\phi|^2, \lb{33z} \ee
where $\phi$ is the Higgs doublet field of the Standard Model. The
Higgs field $\phi$ also interacts directly with field
$\widetilde{\phi}$, according to the interaction (\ref{6mw}) given
by Ref.~\ct{12mw}. It has a time evolution and modifies the shape
of the barrier so that at some value $\widetilde{\phi}_E$ can roll
down the field $\varphi$. This possibility, which we consider in
our paper, is given by the so-called Hybrid Inflation scenarios
\ct{Lin}. Here we assume that the field $\phi$ begins the
inflation at the value $\phi|_{in}\simeq H_0$.

Using the relations given by GWU, we obtain near the local ``false
vacuum'' the following gravitational potential:
\be V(\phi,\,\,\widetilde{\phi}) \simeq \Lambda + \lambda |\phi|^4
+ \widetilde{\lambda}|\widetilde{\phi}|^4 +
a_h|\phi|^2|\widetilde{\phi}|^2, \lb{34z} \ee
where $\lambda=3a_d$  and $\tilde{\lambda} = 3\tilde a_d$ are
self-couplings of the Higgs doublet fields $\phi$ and
$\widetilde{\phi}$, respectively.

\subsection{The GWU action and the self-consistent Higgs Inflation model}

Returning to the problem of the Inflation, we see that the action
of the GWU theory has to be written near the Planck scale as:
\be I_{grav} \simeq - \int_M d^4x \sqrt{-g}
\left(M_{Pl}^{red}\right)^2 \left(\Lambda + \lambda |\phi|^4 +
\tilde \lambda |\widetilde{\phi}|^4 +
a_h|\phi|^2|\widetilde{\phi}|^2
 + ...\right), \lb{35z} \ee
where the cosmological constant $\Lambda$ is almost zero (has an
extremely tiny value).

The next step is to see the evolution of the Inflation in our
model, based on the GWU, with two Higgs fields, $\phi$ and mirror
$\widetilde{\phi}$.

In the present investigation we considered only the result of such
an Inflation, which corresponds to the assumption of the MPP, that
cosmological constant is zero, or almost zero, at both vacua:  at
the "first vacuum" with VEV $v_1 = 246$ GeV and at the "second
vacuum" with VEV $v=v_2\sim 10^{18}$ GeV. If so, we have the
following conditions of the MPP (see section 7):
\be V_{eff} \left(\phi_{min 1}\right) = V_{eff} \left(\phi_{min
2}\right) = 0, \lb{20zz} \ee
\be V'_{eff} \left(\phi_{min 1}\right) = V'_{eff} \left(\phi_{min
2}\right) = 0. \lb{36z} \ee
Considering the total Universe as two worlds, ordinary OW and
mirror MW, we present the following expression for the total
effective Higgs potential, which is far from the Planck scale:

\be V_{eff} = - \mu^2 |\phi|^2 + \lambda (\phi)|\phi|^4 -
\widetilde{\mu}^2 |\widetilde{\phi}|^2 + \widetilde{\lambda}
(\widetilde{\phi}) |\widetilde{\phi}|^4 + \alpha_h (\phi,
\widetilde{\phi}) |\phi|^2 |\widetilde{\phi}|^2,   \lb{37z} \ee
where $\alpha (\phi, \widetilde{\phi})$ is a coupling constant of the
interaction of the ordinary Higgs field $\phi$ with mirror Higgs field
$\widetilde{\phi}$.

According to the MPP, at the critical point of the phase diagram
of our theory, corresponding the "second vacuum", we have: \be \mu
= \widetilde{\mu} = 0, \quad \lambda (\phi_0) \simeq 0, \quad
\widetilde{\lambda}(\widetilde{\phi}_0) \simeq 0, \lb{38z} \ee
and then \be \alpha_h (\phi_0, \widetilde{\phi}_0) = 0,\quad
{\rm{if}}\quad V_{eff}^{crit} (v_2) = 0.  \lb{39z} \ee
At the critical point, corresponding to the first EW vacuum $v_1=246$ GeV, we also
have $V_{eff}^{crit} (v_1) = 0$, according to the MPP prediction of
the existence of degenerate vacua in the Universe.

Then we can present the full scalar Higgs potential by the
following expression:
\be V_{eff} (\phi,\widetilde{\phi}) = \lambda (|\phi|^2 - v^2_1)^2
+ \widetilde{\lambda} (|\widetilde{\phi}|^2 - \widetilde{v}_1^2)^2
+ \alpha_h (\phi, \widetilde{\phi}) (|\widetilde{\phi}|^2 -
\widetilde{v}_1^2) |\phi|^2, \lb{25zz} \ee
where we have shifted the interaction term: \be V_{int} =
\alpha_h(\phi,\widetilde{\phi}) (|\widetilde{\phi}|^2 -
\widetilde{v}^2_1) |\phi|^2   \lb{26zz} \ee
in such a way that the interaction term vanishes, when
$\widetilde{\phi} = \widetilde{\phi}_0 = \widetilde{v}_1$,
recovering the usual Standard Model.

At the end of the Inflation we have: $\widetilde{\phi} =
\widetilde{\phi}_E$, and the first vacuum value of $V_{eff}$ is
given by:
\be V_{eff} (v_1, \widetilde{\phi}_E) =
\widetilde{\lambda}(|\widetilde{\phi}_E|^2 - \widetilde{v}_1^2)^2
 + \alpha_h (v_1,\widetilde{\phi}_E) (|\widetilde{\phi}_E|^2 -
\widetilde{v}_1^2) v_1^2 = 0, \lb{27zz} \ee
and
$$V'_{eff} (v_1,\widetilde{\phi}_E) = \frac{\partial V_{eff}}{\partial
|\phi|^2}
\left|\begin{array}{l}
\phi = v_1\\
\widetilde{\phi} = \widetilde{\phi}_E\end{array}\right|
$$
\be = \alpha_h (v_1,\widetilde{\phi}_E) (|\widetilde{\phi}_E|^2 -
\widetilde{v}^2_1) = 0.
                       \lb{28zz} \ee
This means that the end of the Inflation occurs at the value:
\be \widetilde{\phi}_E = \widetilde{v}_1 = \zeta v_1, \lb{29zz}
\ee
which coincides with the VEV \,$<\widetilde{\phi}>$\,  of the
field $\widetilde{\phi}$ at the first vacuum in the mirror world
MW. Thus,
\be V_{eff} (\phi,\widetilde{\phi}_E) = \lambda (|\phi|^2 -
v^2_1),     \lb{30zz} \ee
which means the Standard Model.

As it is well-known, the total number of e-folds is given by the following expression
(see for example Ref.~\ct{1H}):
\be N^* = \frac{1}{s_{end}^2} \int^{s_{end}}_{s_0}
\frac{ds}{(V'_s/V_s)},  \lb{31zz} \ee
where $s \equiv |\widetilde{\phi}|$ and $s_0=s_{in}$. Then
$$V_s = \lambda (s^2-\widetilde{v}_1^2) + \alpha_h (v_1,s) (s^2 -
\widetilde{v}^2_1) v^2_1,$$ \be V'_s = \frac{\partial
V_s}{\partial s} = 2s \frac{\partial V_s}{\partial s^2}, \lb{32zz}
\ee
and \be \frac{V'_s}{V_s} = 2s
\frac{2\widetilde{\lambda}(s^2-\widetilde{v}^2_1) + \alpha_h
(v_1,s) v^2_1}{\widetilde{\lambda} (s^2-\widetilde{v}^2_1)^2 +
\alpha_h (v_1,s)(s^2-\widetilde{v}^2_1)v^2_1} \,. \lb{33zz} \ee
Then Eq.~(\ref{31z}) gives: \be N^* =
\frac{1}{2s_{end}^2}\int_{s_0}^{s_{end}}
\frac{\lambda(s^2-s_{end}^2)^2 +
\alpha_h(s^2-s_{end}^2)v_1^2}{2\lambda (s^2-s_{end}^2) + \alpha_h
v_1^2}\frac{ds}{s}, \lb{34zz} \ee
where $s=|\widetilde{\phi}|$, $s_{end}=|\widetilde{\phi}_{end}|=\zeta v_1$; $\lambda
= \widetilde{\lambda}$ (property of the MW).\\
As a result, we obtain: \be 8N^* = {(1 +
\frac{\alpha_h}{\lambda}\gamma - 2\gamma)}^2,  \lb{35zz} \ee
where $\gamma = \ln({s_{end}}/{s_0})$ with initial value $s_{in}=s_0$.\\
Using cosmological measurements \ct{3D}, we have:
\be
N^* \simeq 50-60.  \lb{36z} \ee
According to Refs.~\ct{10mw,11mw}, $\zeta \sim 100$, and for
$N^*\simeq 50$ we predict the following estimation:
\be \frac{\alpha_h}{2\lambda} - 1 \simeq \frac{10}{\ln{\zeta} +
\gamma_1}. \lb{37z} \ee
Here $\gamma_1 = \ln(v_1/s_0)$.\\
In cosmology: $s_0\simeq H_0$, where $H_0\simeq 1.5\times {10}^{-42}$ GeV is
the initial Hubble rate, and it is not difficult to estimate that:
\be \frac{\alpha_h}{2\lambda}\sim 1.   \lb{38z} \ee
Finally we obtain: \be \alpha_h(v_1,\widetilde{v}_1)\sim
2\lambda(v_1).  \lb{39z} \ee
Using this information, we conclude that the interaction of the
Higgs and mirror Higgs bosons (see (\ref{6mw})) near the first
vacua is of order of the self-interaction of the ordinary Higgs
bosons near the first EW-vacuum. We also conclude that our theory
really can correspond to $\zeta \sim 100$, as it was estimated  by
Z.~Berezhiani and his collaborators (see Refs.~\ct{10mw,11mw}).

We conclude that the Higgs Inflation scenario developed in this
investigation is self-consistent with our theory based on the
Gravi-Weak Unification, MPM and the discrete space-time Sidharth's
theory at the Planck scale.

Now it is obvious that all previous investigations of the Higgs
field Inflation do not coincide with our model of the Inflation.
It is quite necessary to take into account seriously the
interaction between ordinary and mirror Higgs fields. Thus, it is
not obvious that the Higgs field Inflation from the "false vacuum"
at the Planck scale to the EW vacuum of the Standard Model is in a
disagreement with the MPM predictions of the top-Higgs masses and
modern cosmological parameters' prediction, as it was shown in
Ref.~\ct{1H}, etc.

The theory developed in this investigation predicts the absence of
supersymmetry in the Nature at all, or predicts an essentially
large supersymmetry's breaking scale ($M_{SUSY} > 10^{18}$ GeV),
what is not within the reach of the LHC experiments. We hope that
the future LHC results will shed light on this problem.

In connection with the present investigation, it is necessary to
mention the recent Ref.~\ct{Gor}, in which it was shown that the
RG evolutions of the corresponding Higgs self-interaction
($\lambda(t)$) and Yukawa coupling ($y(t)$) lead to the free-field
stable point:
$$\lambda(M_{Pl})=\dot{\lambda}(M_{Pl})=0$$ in the pure scalar
sector at the Planck scale, what means that the SM is a low-energy
limit of the conceivable theory at the Planck scale, which can be
at least conformal, or even superconformal one. This circumstance
is phenomenologically motivated by the actual properties of the
SM. In this scenario, the Higgs sector could emerge as a Goldstone
boson, associated with a spontaneous breaking of the high-energy
conformal invariance. This would simultaneously resolve the
hierarchy and Landau pole problems in the scalar sector and would
provide a nearly flat potential with two almost degenerate vacua
at the EW and Planck scale.

\section{Summary and Conclusions}

\begin{enumerate}

\item We suggested to consider the theory of a discrete space-time
by B.G. Sidharth as a theory of the Planck scale physics, existing
at the early stage of the evolution of our Universe. We have used
the Sidharth's predictions of the  non-commutativity to have
almost zero cosmological constant (c.c.). Previously B.G.Sidharth
was first who has shown that c.c. is $\Lambda \sim H_0^2$, where
$H_0$ is the Hubble rate, and the Dark Energy density is very
small ($\sim 10^{-48}\,\,{\rm{GeV}}^4$), what provided the
accelerating expansion of our Universe after the Big Bang. This
result of almost zero c.c. was applied to our Gravi-Weak
Unification (GWU) model.
\item Using the Plebanski's formulation of gravity, we constructed
the Gravi-Weak Unification model, which is invariant under the
$G_{(GWU)} = Spin(4,4)$-group, isomorphic to the $SO(4,4)$-group.
Gravi-Weak Unification is a model unifying gravity with the weak
$SU(2)$ gauge and Higgs fields.
\item  We considered also the ideas of the Random Dynamics,
developed by H.B. Nielsen and his collaborators, with aim to
explain, why the Nature has chosen at the early stage the symmetry
$G = G_{(GWU)}\times U(4)$, where $U(4)$ is a group of fermions.
Random Dynamics also assumes a discrete space-time (with
lattice-like structure), and leads to the Multiple Point Principle
(MPP), which postulates that the Nature has the Multiple Critical
Point (MCP). The MPP-model (MPM) predicts the existence of several
degenerate vacua in the Universe, all having zero or almost zero
cosmological constants.
\item In contrast to other theories of unification, we accepted an
assumption of the existence of visible and invisible (hidden)
sectors of the Universe. We gave arguments that modern
astrophysical and cosmological measurements lead to a model of the
Mirror World with a broken Mirror Parity (MP), in which the Higgs
VEVs of the visible and invisible worlds are not equal:
$\langle\phi\rangle=v, \quad
\langle\widetilde{\phi}\rangle=\widetilde{v} \quad {\rm{and}}\quad
v\neq \widetilde{v}$. We considered a parameter characterizing the
violation of the MP: $\zeta = \widetilde{v}/v \gg 1$, using the
result: $\zeta \sim 100$ obtained by Z.~Berezhiani and his
collaborators.
\item In our model we showed that the action for gravitational and
$SU(2)$ Yang--Mills and Higgs fields, constructed in the ordinary
world (OW), has a modified duplication for the hidden (mirror)
world (MW) of the Universe.
\item Considering the Gravi-Weak symmetry breaking, we have
obtained the following sub-algebras: $\mathfrak{g_1} =\mathfrak
{sl}(2,C)^{(grav)}_L \oplus \mathfrak{su}(2)_L$ -- in the ordinary
world, and $\mathfrak{\widetilde{g_1}} =
\mathfrak{\widetilde{sl}}(2,C)_R^{(grav)} \oplus \mathfrak
{\widetilde{su}}(2)_R$ -- in the hidden world. These sub-algebras
contain the self-dual left-handed gravity in the OW, and the
anti-self-dual right-handed gravity in the MW. We showed, that
finally at low energies we have the Standard Model and the
Einstein-Hilbert's gravity.
\item We reviewed the Multiple Point Model (MPM) by D.L. Bennett
and H.B.Nielsen. We showed that the existence of two vacua into
the SM: the first one -- at the Electroweak scale ($v_1\simeq 246$
GeV), and the second one -- at the Planck scale ($v_2\sim 10^{18}$
GeV), was confirmed by calculations of the Higgs effective
potential in the 2-loop and 3-loop approximations. The
Froggatt-Nielsen's prediction of the top-quark and Higgs masses
was given in the assumption that there exist two degenerate vacua
into the SM. It was shown that this prediction was improved by the
next order calculations.
\item We have developed a model of the Higgs Inflation using the
GWU action, which contains a non-minimal coupling of the Higgs
field with gravity, suggested by F. Bezrukov and M. Shaposhnikov.
According to this model, a scalar field $\sigma$, being an
inflaton, starts trapped from the "false vacuum" of the Universe
at the value of the Higgs field's VEV $v =v_2 \sim 10^{18}$ GeV.
Then during the Inflation $\sigma$ decays into the two Higgs
doublets of the SM: $\sigma\to \phi^\dagger \phi$. The interaction
between the ordinary and mirror Higgs fields $\phi$ and
$\widetilde{\phi}$ generates a Hybrid model of the Higgs Inflation
in the Universe. Such an interaction leads to the emergence of the
SM vacua at the EW scales: with the Higgs boson VEVs $v_1\approx
246$ GeV -- in the OW, and $\widetilde{v}_1=\zeta v_1$ -- in the
MW. Our model of the Higgs Inflation is in agreement with the
predictions of the top-Higgs masses, $\zeta \sim 100$ and modern
cosmological parameters $N^*,\,A_s$ and $r$.
\item The GWU theory developed in this investigation predicts the
absence of the supersymmetry at least before $10^{18}$ GeV.

\end{enumerate}

\section{Acknowledgments}

We thank Prof. M.~Chaichian for useful discussions, and Dr
C.R.~Das for his help. L.V.L. greatly thanks the B.M.~Birla
Science Centre (Hyderabad, India) and personally Prof.
B.G.~Sidharth, for hospitality, collaboration and financial
support. H.B.N. wishes to thank the Niels Bohr Institute for the
status of professor emeritus and corresponding support.


\begin{thebibliography}{99}
\bibitem{1S} B.G.~Sidharth,  The Thermodynamic Universe. World Scientific,
Singapore, 2008.
\bibitem{2S} B.G.~Sidharth, Int.J. of Mod. Phys. A{\bf 13}, 2599 (1998).
\bibitem{3S} B.G.~Sidharth, {\it  Proc. of the 8th Marcell Grossmann Meeting on
General Relativity}, Jerusalem, May 1997. Ed. T. Piran, World
Scientific, Singapore, 1999, p.476.
\bibitem{4S} B.G.~Sidharth, Found.Phys. {\bf 38} (1), 89 (2008);
ibid., {\bf 38} (8), 695 (2008); Int. J. Th. Phys. {\bf 37}, 1307
(1998); ibid., {\bf 43} (9), 1857 (2004);  arXiv:1201.0915.
\bibitem{1R} H.B.~Nielsen, D.L.~Bennett and N.~Brene, {\it The random dynamics
project or from fundamental to human physics}, Recent Developments
in Quantum Field Theory (1985). Ed. J.~Ambjorn, B.J.~Durhuus,
J.L.~Petersen, ISBN 0444869786.
\bibitem{2R} C.D.~Froggatt and H.B.~Nielsen, {\it Origin of Symmetries}, Singapore:
World Scientific, 1991.
\bibitem{3R} H.B.~Nielsen and N.~Brene, {\it Some remarks on random dynamics},
Proof the 2nd Nishinomya Yukawa Memorial Symposium on String
Theory, Kyoto University, 1987 (Sprinter, Berlin, 1988); Phys.
Lett. B{\bf 233}, 399 (1989);  Nucl. Phys. B{\bf 359}, 406 (1991).
\bibitem{4R} H.B.~Nielsen, {\it Random Dynamics and relations between the number
of fermion generations and fine structure constant}. A talk
presented at Zakopane Summer School, May 11-Jun 10, 1988; Acta
Physica Polonica, B{\bf 20}, 427 (1989).
\bibitem{5R} H.B.~Nielsen, S.E.~Rugh, Nucl. Phys. C{\bf 29}, 200 (1992).
\bibitem{6R} H.B.~Nielsen and A.~Kleppe, in: {\it Proceedings to the 16th Workshop
on "What Comes Beyond the Standard Models?"\,}, Slovenia, Bled,
14-21 July, 2014, Vol.14, No.2, DMFA Zaloznistvo, Ljubljana,
arXiv:1403.14101.
\bibitem{7R} H.B.~Nielsen and M.~Ninomiya, in:
{\it Proceedings to the 9th Workshop on "What Comes Beyond the
Standard Models?"\,}, Bled, Slovenia, September 16-26, 2006, DMFA
Zaloznistvo, Ljubljana, hep-ph/0612032; in: {\it Proceedings to
the 13th Workshop on "What Comes Beyond the Standard Models?"\,},
Bled, Slovenia, July 12-22, 2010, DMFA Zaloznistvo, Ljubljana,
arXiv:1008.0464.
\bibitem{8R} L.V. Laperashvili, {\it The Standard Model and the fine
structure constant at the Planck distances in
Bennett-Brene-Nielsen-Picek random dynamics}, Phys. Atom. Nucl.
{\bf 57}, 471 (1994), [Yad. Fiz. {\bf 57}, 501 (1994)].
\bibitem{1gw}
S.~Alexander, {\it Isogravity: Toward an Electroweak and
Gravitational Unification}, arXiv:0706.4481.
\bibitem{2gw}
F.~Nesti, Eur. Phys. J. C {\bf 59}, 723 (2009), arXiv:0706.3304.
\bibitem{3gw}
S.~Alexander, A.~Marciano and L.~Smolin, {\it Gravitational origin
of the weak interaction's chirality}, arXiv:1212.5246.
\bibitem{1Lizi} A.~Garrett~Lisi, {\it An Exceptionally Simple Theory of
the Everything}, arXiv:0711.0770; J. Phys. A{\bf 43}, 445401
(2010), arXiv:1004.4866.
\bibitem{1gu}
F.~Nesti and R.~Percacci, Phys. Rev. D {\bf 81}, 025010 (2010),
arXiv:0909.4537.
\bibitem{2gu}
A.~Torres-Gomez and K.~Krasnov, Phys. Rev. D {\bf 81}, 085003
(2010), arXiv:0911.3793.
\bibitem{3gu}
L.~Smolin, Phys. Rev. D {\bf 80}, 124017 (2009), arXiv:0712.0977.
\bibitem{1u} C.R.~Das, L.V.~Laperashvili and A.~Tureanu, Int. J. Mod. Phys. A{\bf
28}, 1350085 (2013), arXiv:1304.3069.
\bibitem{2u} A.~Garrett Lisi, L.~Smolin and S.~Speziale, J. Phys. A{\bf 43},
445401 (2010), arXiv:1004.4866.
\bibitem{1mw} R.~Foot, H.~Lew and R.R.~Volkas, Phys. Lett. B{\bf 272}, 67 (1991).
\bibitem{2mw} R.~Foot, Mod. Phys. Lett. A{\bf 9}, 169 (1994), arXiv:hep-ph/9402241.
\bibitem{3mw} Z.~Berezhiani, A.~Dolgov and R.N.~Mohapatra, Phys. Lett. B{\bf 375}, 26 (1996), arXiv:hep-ph/9511221.
\bibitem{4mw} R.~Foot, Int. J. Mod. Phys. D{\bf 13}, 2161 (2004), arXiv:astro-ph/0407623.
\bibitem{5mw} Z.~Berezhiani,{\it Through the looking-glass: Alice's adventures in
mirror world}, in: Ian Kogan Memorial Collection "From Fields to
Strings: Circumnavigating Theoretical Physics", Vol. 3, eds. M.
Shifman et al. (World Scientific, Singapore, 2005), pp. 2147-2195,
arXiv:hep-ph/0508233.
\bibitem{6mw} L.B.~Okun, Phys. Usp. {\bf 50}, 380 (2007),
arXiv:hep-ph/0606202.
\bibitem{7mw} S.I.~Blinnikov, Phys. Atom. Nucl. {\bf 73},
593 (2010), arXiv:0904.3609.
\bibitem{8mw} P.~Ciarcelluti, Int. J. Mod. Phys. D{\bf 19}, 2151
(2010), arXiv:1102.5530.
\bibitem{9mw} J.-W.~Cui, H.-J.~He, L.-C.~Lu and F.-R.~Yin, Phys. Rev. D{\bf 85}, 096003 (2012), arXiv:1110.6893 [hep-ph].
\bibitem{Sil} Z.K.~Silagadze, {\it Mirror dark matter discovered?}, ICFAI
U. J. Phys. {\bf 2}, 143 (2009), arXiv:0808.2595.
\bibitem{Lin} A.~Linde, {\it Hybrid Inflation}, Phys. Rev. D{\bf 49},
748 (1994), arXiv:astro-ph/9307002.
\bibitem{Rov} C.~Rovelli, {\it Quantum Gravity}, Cambridge Monographs
on Mathematical Physics, 2007.
\bibitem{1N} L.~Nottale, Int. J. Mod. Phys. A{\bf 7}, 4899 (1992);
{\it Fractal Space-Time and Microphysics}, World Scientific,
Singapore, 1993; {\it Frontiers of Fundamental Physics},
Proceeding of Fourth International Symposium, Hyderabad, India,
11-13 December, 2000. Ed. B.G. Sidharth.
\bibitem{1k} H.~Kleinert, {\it Multivalued Fields in Condensed Matter,
Electromagnetism, and Gravitation}, World Scientific, Singapore,
2008; EJTP {\bf 7}, 287 (2010).
\bibitem{3k} D.L.~Bennett, C.R.~Das, L.V.~Laperashvili and
H.B.~Nielsen, Int. J. Mod. Phys. A{\bf 28}, 1350044 (2013).
\bibitem{Per} S.~Perlmutter et al., Astrophys.J. {\bf 517}, 565
(1999), arXiv:astro-ph/9812133.
\bibitem{1F} H.S.~Snyder, Phys. Rev. {\bf 72}, 68 (1947).
\bibitem{2F} D.~Amati, {\it Sakharov Memorial Lectures}. Ed.Kaddysh,
L.V. and Feinberg, N.Y., Nova Science, N.Y., p.455.
\bibitem{3F} A.~Kempf, {\it From the Planck Length to the Hubble Radius}.
Ed. A.~Zichichi, World Scientific, Singapore, 1995, p.613.
\bibitem{4F} J.~Madore, Class. Quant. Grav., {\bf 9}, 69 (1992).
\bibitem{10S} B.G.~Sidharth, {\it The Chaotic Universe: From the Planck
to the Hubble Scale}, Nova Science, New York (2001).
\bibitem{1Z} Ya.B.~Zeldovich, JETP Lett. {\bf 6}, 316 (1967).
\bibitem{1W} S.~Weinberg, Phys. Rev. Lett. {\bf 43}, 1566 (1979).
\bibitem{1D} A.~Riess et al., Astrophys. J. Suppl. {\bf 183}, 109
(2009), arXiv:0905.0697.
\bibitem{2D} W.L.~Freedman et al., Astrophys. J. {\bf 704}, 1036
(2009), arXiv:0907.4524.
\bibitem{3D} K.A.~Olive  et al. (Particle Data Group), Chin. Phys. C{\bf 38},
090001 (2014).
\bibitem{11S} B.G.~Sidharth, Found. Phys. Lett. {\bf 18}(4), 393
(2005); ibid., {\bf 18}(7), 757 (2006).
\bibitem{CNT} M.~Chaichian, K.~Nishijima and A.~Tureanu, {\it Spin
Statistics and CPT Theorems in Non-Commutative Field Theory},
Phys. Lett. B{\bf 568}, 146 (2003), arXiv:hep-th/0209008.
\bibitem {1M} D.L.~Bennett and H.B.~Nielsen, Int. J. Mod. Phys. A{\bf 9}, 5155
(1994).
\bibitem {2M} C.D.~Froggatt and H.B.~Nielsen, Phys. Lett. B{\bf 368}, 96 (1996),
arXiv:hep-ph/9511371.
\bibitem{1AG}
C.D.~Froggatt, G.~Lowe, H.B.~Nielsen, {\it Antigrand unification
and the fermion mass problem}, Nucl. Phys. B{\bf 414}, 579 (1994).
\bibitem{2AG}
L.V.~Laperashvili, {\it Antigrand unification and the phase
transitions at the Planck scale in gauge theories}. Invited talk
at the Fourth International Symposium on Frontiers of Fundamental
Physics, 11-13 Dec 2000, Hyderabad, India, in: Proceeding of the
Fourth International Symposium, 2000. Ed. B.G. Sidharth,
arXiv:hep-th/0101230.
\bibitem{1A-H} N.~Arkani-Hamed, {\it Invited talk at the Conference on
Hierarchy Problems in Four and More Dimensions}, ICTP, Italy,
Trieste, 1-4 Oct, 2003.
\bibitem{1V} G.~Volovik, JETP Lett. {\bf 79}, 101 (2004).
\bibitem{3P} J.F.~Plebanski, J. Math. Phys. {\bf 18}, 2511 (1977).
\bibitem{4P} A.~Ashtekar, Phys. Rev. Lett. {\bf 57}, 2244 (1986);
Phys. Rev. D{\bf 36}, 1587 (1987).
\bibitem{6P} R.~Capovilla, T.~Jacobson, J.~Dell and L.J.~Mason,
Class. Quant. Grav. {\bf 8}, 41 (1991).
\bibitem{7P} R.~Capovilla, T.~Jacobson and J.~Dell,
Class. Quant. Grav. {\bf 8}, 59 (1991).
\bibitem{3} D.L.~Bennett, L.V.~Laperashvili, H.B.~Nielsen and A.~Tureanu,
Int. J. Mod. Phys. A{\bf 28}, 1350035 (2013), arXiv:1206.3497.
\bibitem{4} L.V.~Laperashvili, H.B.~Nielsen, A.~Tureanu, arXiv:1411.6456,
to be published in  Int. J. Mod. Phys. D (2015).
\bibitem{3u} E.W.~Mielke, Phys. Rev. D{\bf 77}, 084020 (2008), arXiv:0707.3466.
\bibitem{3uu} G. de Berredo-Peixoto and I.L.~Shapiro,
Phys. Rev. D{\bf 70}, 044024 (2004), arXiv:hep-th/0307030.
\bibitem{4u} D.L.~Bennett, L.V.~Laperashvili and H.B.~Nielsen,
{\it Relation between finestructure constants at the Planck scale
from multiple point principle}, in: Proceedings to the 9th
Workshop on "What Comes Beyond the Standard Models?'\,, Bled,
Slovenia, July 16-27, 2006 (DMFA, Zaloznistvo, Ljubljana, 2006),
arXiv:hep-ph/0612250; {\it Finestructure constants at the Planck
scale from multiple point principle}, in: Proceedings to the 10th
Workshop on "What Comes Beyond the Standard Models?"\,, Bled,
Slovenia, July 17-27, 2007 (DMFA, Zaloznistvo, Ljubljana, 2007),
arXiv:0711.4681.
\bibitem{LY} T.D.~Lee and C.N.~Yang, Phys. Rev. {\bf 104}, 254 (1956).
\bibitem{KOP} I.Yu.~Kobzarev, L.B.~Okun and I.Ya.~Pomeranchuk,
Yad. Fiz. {\bf 3}, 1154 (1966) [Sov. J. Nucl. Phys. {\bf 3}, 837
(1966)].
\bibitem{10mw} E.K.~Akhmedov, Z.G.~Berezhiani and G.~Senjanovic,
Phys. Rev. Lett. {\bf 69}, 3013 (1992), arXiv:hep-ph/9205230.
\bibitem{11mw} Z.~Berezhiani, P.~Ciarcelluti, D.~Comelli and F.L.~Villante,
Int. J. Mod. Phys. D{\bf 14}, 107 (2005), arXiv:astro-ph/0312605.
\bibitem{12mw} R.~Foot, A.~Kobakhidze and R.R.~Volkas, Phys. Rev.
D{\bf 84}, 09503 (2011), arXiv:1109.0919.
\bibitem{1cl} F.~Nesti, Eur. Phys. J. C{\bf 59}, 723 (2009), arXiv:0706.3304.
\bibitem{1Hig} G.~Aad et al. [ATLAS Collaboration], Phys. Lett. B{\bf 716},
1 (2012), arXiv:1207.7214.
\bibitem{2Hig} S.~Chatrchyan et al. [CMS Collaboration], Phys. Lett. B{\bf 716}, 30 (2012), arXiv:1207.7235.
\bibitem{1nbs} C.D.~Froggatt and H.B.~Nielsen, {\it Surveys High Energy
Phys.}, {\bf 18}, 55 (2003), arXiv:hep-ph/0308144.
\bibitem{2nbs} C.D ~Froggatt, H.B.~Nielsen and L.V.~Laperashvili,
Int. J. Mod. Phys. A{\bf 20}, 1268 (2005), arXiv:hep-ph/0406110.
\bibitem{3nbs} C.D.~Froggatt, {\it PASCOS 2004 Particles, Strings and
Cosmology}, ed. by G.~Alverson, E.~Barberis, P.~Nath and
M.T.~Vaughn, World Scientific, Singapore, 2005, pp.325-334,
arXiv:hep-ph/0412337.
\bibitem{9} C.D.~Froggatt, L.V.~Laperashvili and H.B.~Nielsen,
Phys. Atom. Nucl. {\bf 69}, 67 (2006) [Yad. Fiz. {\bf 69}, 3
(2006)], arXiv:hep-ph/0407102.
\bibitem{26Fr}
N.~Cabbibo, L.~Maiani, G.~Parisi and R.~Petronzio, Nucl. Phys.
B{\bf 158}, 295 (1979).
\bibitem{27Fr} P.Q.~Hung, Phys. Rev. Lett. {\bf 42}, 873 (1979).
\bibitem{28Fr} M.~Lindner, Z. Phys. C{\bf 31}, 295 (1986).
\bibitem{29Fr} M.~Sher, Phys. Lett. B{\bf 317}, 159 (1993);
Addendum: Phys. Lett. B{\bf 331}, 448 (1994).
\bibitem{30Fr}
G.~Altarelli and G.~Isidori, Phys. Lett. B{\bf 337}, 141 (1994).
\bibitem{31Fr}
J.A.~Casas, J.R.~Espinosa and M.~Quiros, Phys. Lett. B{\bf 342},
171 (1995).
\bibitem{32Fr} J.~Elias-Miro, J.~R.~Espinosa, G.~F.~Giudice, G.~Isidori,
A.~Riotto and A.~Strumia, Phys. Lett. B{\bf 709}, 222 (2012),
arXiv:1112.3022.
\bibitem{33Fr}
F.~Bezrukov, M.Y.~Kalmykov, B.A.~Kniehl and M.~Shaposhnikov, JHEP
{\bf 1210}, 140 (2012), arXiv:1205.2893.
\bibitem{34Fr}
G.~Degrassi, S.~Di~Vita, J.~Elias-Miro, J.R.~Espinosa,
G.F.~Giudice, G.~Isidori, A.~Strumia, JHEP {\bf 1208}, 098
(2012), arXiv:1205.6497.
\bibitem{35Fr}
D.~Buttazzo, G.~Degrassi, P.~P.~Giardino, G.~F.~Giudice, F.~Sala,
A.~Salvio and A.~Strumia, JHEP {\bf 12}, 089 (2013),
arXiv:1307.3536.
\bibitem{FNT} C.D.~Froggatt, H.B.~Nielsen and Y.~Takanishi, Phys.
Rev. D{\bf 64}, 113014 (2001), arXiv:hep-ph/0104161.
\bibitem{4nbs} C.D.~Froggatt and H.B.~Nielsen, Phys. Rev. D {\bf 80}, 034033
(2009), arXiv:0811.2089.
\bibitem{5nbs} C.R.~Das, L.V.~Laperashvili, {\it Phase transition
in gauge theories, monopoles and the multiple point principle},
Int. J. Mod. Phys. A{\bf 20}, 5911 (2005), arXiv:hep-ph/0503138.
\bibitem{6nbs} C.D.~Froggatt and H.B.~Nielsen, {\it Tunguska Dark Matter Ball},
arXiv:1403.7177.
\bibitem{1Br} V. Brancina, E. Messina and M. Sher, {\it The lifetime
of the electroweak vacuum and sensitivity to Planck scale physics},
arXiv:1408.5302.
\bibitem{2Br}  V.~Brancina, E.~Messina and A.~Platania, JHEP {\bf 1409},
182 (2014), arXiv:1407.4112.
\bibitem{3Br} V.~Brancina and E.~Messina, Phys. Rev. Lett. {\bf 111},
241801 (2013), arXiv:1307.5193.
\bibitem{1Bez} F.L.~Bezrukov and M.~Shaposhnikov,
Phys. Lett. B{\bf 659}, 703 (2008), arXiv:0710.3755,
arXiv:1411.1923.
\bibitem{2Bez} F.L.~Bezrukov and  Gorbunov,
JHEP {\bf 1307}, 140 (2013), arXiv:1303.4395.
\bibitem{3Bez}
F.~Bezrukov, J.~Rubio and M.~Shaposhnikov, {\it Living beyond the
edge: Higgs inflation and vacuum metastability}, arXiv:1412.3811.
\bibitem{1MT} I.~Masina and A.~Notari, Phys. Rev. Lett. {\bf 108},
191302 (2012), arXiv:1112.5430; JCAP {\bf 1211}, 031 (2012),
arXiv:1204.4155.
\bibitem{1H}  M.~Fairbairn, P.~Grothus and R.~Hogan, JCAP {\bf 1406},
039 (2014), arXiv:1403.7483.
\bibitem{2H} M.~Fairbairn, R.~Hogan. Phys. Rev. Lett. {\bf 112}, 201801
(2014), arXiv:1403.6786.
\bibitem{Gor}
A.~Gorsky, A.~Mironov, A.~Morozov and T.N.~Tomaras, {\it Is the
Standard Model saved asymptotically by conformal symmetry?}
FIAN-TD-12-14, ITEP-TH-24-14, arXiv:1409.0492.


\end{thebibliography}
\end{document}